\documentclass{article}
\usepackage[margin=1in]{geometry}
\usepackage{graphicx} 
\usepackage{amsmath}
\usepackage{enumerate}
\usepackage{natbib}
\usepackage{amsmath, calc}
\usepackage{booktabs}
\usepackage{multirow}
\usepackage{rotating, tabularx}
\usepackage{tikz}
\usepackage{url} 
\usepackage{subcaption}
\usepackage{float}
\usepackage{authblk}

\title{Time-dependent Predictive Accuracy Metrics in the Context of Interval Censoring and Competing Risks}

\author[1]{Zhenwei Yang}
\author[1]{Dimitris Rizopoulos}
\author[2]{Lisa F. Newcomb}
\author[1,3]{Nicole S. Erler}
\affil[1]{Department of Epidemiology and Biostatistics, Erasmus Medical Center Rotterdam}
\affil[2]{Fred Hutchinson Cancer Center, Cancer Prevention Program, Public Health Sciences, Seattle, Washington}
\affil[3]{Julius Center for Health Sciences and Primary Care, University Medical Center Utrecht, Utrecht University}

\date{\vspace{-5ex}}

\begin{document}

\maketitle

\begin{abstract}
Evaluating the performance of a prediction model is a common task in medical statistics. Standard accuracy metrics require the observation of the true outcomes. This is typically not possible in the setting with time-to-event outcomes due to censoring. Interval censoring, the presence of time-varying covariates, and competing risks present additional challenges in obtaining those accuracy metrics. In this study, we propose two methods to deal with interval censoring in a time-varying competing risk setting: a model-based approach and the inverse probability of censoring weighting (IPCW) approach, focusing on three key time-dependent metrics: area under the receiver-operating characteristic curve (AUC), Brier score, and expected predictive cross-entropy (EPCE). The evaluation is conducted over a medically relevant time interval of interest, $[t, \Delta t)$. The model-based approach includes all subjects in the risk set, using their predicted risks to contribute to the accuracy metrics. In contrast, the IPCW approach only considers the subset of subjects who are known to be event-free or experience the event within the interval of interest. We performed a simulation study to compare the performance of the two approaches with regard to the three metrics. Furthermore, we demonstrated the three metrics using the two approaches on an example prostate cancer surveillance cohort. Risk predictions were generated from a joint model handling the interval-censored cancer progression and the competing event, early treatment, and repeatedly measured biomarkers.
\end{abstract}


\maketitle

\section{Introduction}

Accurate evaluation of a model's predictive accuracy is essential for developing reliable prediction models to guide clinical decisions and public health strategies. Commonly used metrics to evaluate prediction models include the Brier score (to determine how close predictions are to the actual outcomes)\citep{Brier1950},  the area under the receiver operating characteristic (ROC) curve, AUC (to distinguish between low- and high-risk individuals)\citep{Zweig1993}, and the expected predictive cross-entropy (EPCE, to compare the predicted probabilities of events happening and the actual observed events)\citep{Commenges2012,commenges2015}. Calculation of these requires researchers to observe the actual outcome.\citep{Steyerberg2010} This is usually the case for continuous and dichotomous outcomes. However, for time-to-event data, censoring is common and poses difficulty in determining the occurrence of the outcomes.\citep{Leung1997} The most common type, right censoring, has been extensively investigated. For example, \citet{Heagerty2000} and \citet{Wang2009} proposed to estimate the AUC via the Kaplan-Meier estimator or an estimator of the bivariate distribution function of the actual outcome and the event time variable. \citet{Heagerty2005} demonstrated the calculation of the AUC based on the survival regression models. Harrell's concordance index, as a discrimination metric, deals with right censoring by excluding the patients censored before the time of interest.\citep{Harrell1996} 

In settings with interval censoring, the situation is even more complicated because also for patients who are known to have experienced the event of interest during the follow-up, the exact event time is unknown.\citep{Zhang2010} Consequently, the ambiguity of the underlying unobserved event times bring challenges in defining cases (patients who experience the event before a timepoint or within a particular time interval) and controls (patients who ``survived" up to a timepoint or time interval). The identification of cases and controls is essential in formulating the sensitivity and specificity of the prediction model, based on which the ROC curve is calculated. \citet{Tsouprou2015} described a method to order the event times of any two interval-censored subjects via a joint density of the two estimated event times when calculating the concordance index. Time-varying covariates have been drawing greater attention recently in medical applications. The evaluation then needs to be done for a specific time point or a time interval, and the metrics should be a function of time. To accommodate this, time-dependent evaluation metrics were developed, which allow for the dynamic determination of cases and controls.\citep{Bansal2018} Researchers can then evaluate the model using its predicted risk for an event before a time $t + \Delta t$ conditional on covariate information and being event-free up until a time $t$. This can be seen as the risk of experiencing an event in a clinically relevant interval $[t, t+dt)$. Last but not least, another common complication is the presence of competing events: patients who are observed with the competing event might still have experienced the (yet unobserved) interval-censored event. This combination of competing risk and interval censoring is not a rare occurrence in clinical research, for instance, in oncology. Cancer progression or diagnosis is typically not directly observable and unless all-cause mortality is the event of interest, death poses a competing risk. This has to be taken into account when obtaining a risk estimate but also further complicates the determination of whether the patient is a case or control during model evaluation. Although others have proposed adaptions of the definition of sensitivity being event-specific,\citep{Saha2010, Dey2020} the combination of the abovementioned difficulties brings forth challenges in the model evaluation.


One notable study by \citet{Hélène2016} investigated three approaches to calculate the AUC for a multi-state model with competing risks and interval censoring, but only considered baseline covariates. The authors proposed two model-based approaches in which subjects contribute to the sensitivity and specificity in a weighted manner depending on their predicted risk from the model. The other approach corrects for the censoring in a nonparametric manner, which uses the inverse probability of censoring weighting (IPCW), whereas the risk probabilities are still derived from a parametric model. \citet{Hélène2016} concentrated on marker discrimination, whereas clinical decision-making often relies on model discrimination.\citep{Leening2014} Except for the AUC as an example of the predictive accuracy metrics, \citet{Blanche2014} proposed the estimation of the Brier score using the IPCW approach in the setting with competing events and time-varying covariates. They did not, however, consider the setting with interval censoring. \citet{Rizopoulos2024} extended the idea of the EPCE proposed by \citet{Commenges2012} and used that as an evaluation metric for a medical-relevant time interval in the setting only with time-varying covariates. 

This work aims to extend the formulation of time-dependent predictive accuracy metrics, specifically the AUC, Brier score and EPCE, in the context of interval-censored data with competing risks. We present the adapted accuracy metrics using the Canary Prostate Active Surveillance Study (PASS).\citep{Newcomb2016} Patients with low-risk prostate cancer were admitted to active surveillance instead of receiving immediate and typically invasive treatment, and closely monitored by regular biopsies and biomarker measurements (prostate-specific antigen, PSA). Only when a progression of the cancer is diagnosed, treatment is recommended for these patients. The primary event of interest is prostate cancer progression, which is interval-censored between a negative biopsy and a positive biopsy. Since the start of treatment means the end of active surveillance, cancer progression while on active surveillance can no longer be observed, thus the treatment initiation constitutes a competing risk. The example prediction model, based on the motivating data, is a joint model (ICJM) for repeatedly measured PSA and the time to both the primary and competing events.\citep{Yang2023}  We explore and extend two methods to evaluate the predictive performance in the context of interval censoring and competing risk: a model-based approach and the IPCW approach. The former takes advantage of all patients at risk at the beginning of an interval of interest, $[t, t+dt)$, to determine the probabilities of being a case or control for the particular interval and weighs those patients based on their predicted risks from the model or algorithm itself. The latter uses only the subset of the sample for which it is known whether they are a case or a control, and weights this subset to represent the whole sample. We compare the AUC and Brier scores using both approaches and explore the impact of model misspecification as well as different patterns of interval censoring. For the model-based approach, we also evaluate the EPCE. The corresponding source codes can be found on GitHub (https://github.com/ZhenweiYang96/Predicitve-accuracy-metrics). The evaluation metrics are applicable to general statistical models as well as other prediction algorithms such as machine learning techniques.


The remainder of this paper is presented as follows. The detailed methodology of estimating the three time-dependent accuracy metrics is demonstrated in Section~\ref{sec:method}. In Section~\ref{sec:app}, we present the application of these methods to the interval-censored cause-specific joint model (ICJM). The results from two simulation studies to investigate the two approaches used on different metrics are displayed in Section~\ref{sec:sim}. At last, there follows a discussion part in Section~\ref{sec:disc}.

\section{Predictive accuracy metrics} \label{sec:method}


\subsection{Notation}
Motivated by the Canary PASS data, we denote $\delta$ as the observed event type of patients in the test set, with $\delta = 1$ being cancer progression, $\delta = 2$ early treatment, and $\delta = 3$ right censoring  (e.g., due to loss to follow-up or at the time of data extraction). We distinguish between the true (but unobserved) time of cancer progression ($T^{\textsc{prg}*}$), the true time of early treatment initiation ($T^{\textsc{trt}*}$), and the censoring time ($T^{\textsc{cen}}$). For patients recorded with the interval-censored event, cancer progression ($T^{\textsc{prg}*}<\min\left\{T^{\textsc{cen}}, T^{\textsc{trt}*}\right\}$), it is only known that progression happened between the last progression-free biopsy (i.e., the observed progression-free time), $T^{\textsc{prg-}}$, and the biopsy time at which progression was detected, $T^{\textsc{prg+}}$. For the patients who started early treatment ($T^{\textsc{trt}*}<\min\left\{T^{\textsc{cen}}, T^{\textsc{prg+}}\right\}$), progression still might have occurred in the interval between the last progression-free biopsy and the treatment initiation time $T^{\textsc{trt}} = T^{\textsc{trt}*}$ but was not detected. For patients who are censored at $T^{\textsc{cen}}$, the corresponding progression-risk interval is between the last progression-free biopsy and infinity.

We denote the vector of repeated observations of the PSA level with $\boldsymbol{\mathcal{Y}}$. We distinguish the training dataset, used to fit the prediction model, and the test dataset, in which the prediction model is being evaluated. The training set is denoted by $\boldsymbol{\mathcal{D}}_n$ and includes the longitudinal PSA measurements, event times and other (baseline) covariates of $n$ subjects. We also define an interval of interest for prediction, $[t, t + \Delta t)$. The quantity of interest is the progression-specific risk at $t+\Delta t$ conditional on no event occurring before time $t$, denoted by $\Pi^{\textsc{prg}}(t+\Delta t\mid t)$ and can be formulated as
\begin{align*}
\Pi^{\textsc{prg}}(t+\Delta t\mid t) = \Pr\Bigg[T^{\textsc{prg}*} \leq t + \Delta t, T^{\textsc{prg}*} < T^{\textsc{trt}*} \mid T^{\textsc{prg}*} > t, T^{\textsc{trt}*} > t, \boldsymbol{\mathcal{X}}(t), \boldsymbol{\mathcal{D}}_n\Bigg],
\end{align*}
where $\boldsymbol{\mathcal{X}}(t)$ denotes the data of the test set, including the longitudinal PSA measurements up until time $t$ and all other (baseline) covariates used in the prediction. Typically, a subject is considered as a case if their event occurs within the interval of interest, and as a control if the event is after that interval. In the test set, the $n_t$ patients who have not progressed, been treated, or censored before time $t$ form the risk set and are considered in calculating the evaluation metrics in the following subsections.

\subsection{Definition of cases and controls} \label{sec:dcc}

The definition and calculation of the AUC and Brier score stems from classifying subjects into cases and controls, i.e., whether the subject experiences the event during an interval of interest. Contrary to the case of right-censored data, for an interval-censored event, the event time is presented by a risk interval, rather than a single time point. The relative positions of the risk interval and the interval of interest (see Figure~\ref{fig:relpos}) bring challenges in clearly distinguishing patients being a case or control. The risk interval overlapping with the interval of interest means that the patient has a certain probability of being a case. When the risk interval contains values larger than $t + \Delta t$, the patient has a particular probability of being a control.

\begin{figure}[H]
    \centering
    \tikzset{every picture/.style={line width=0.75pt}} 
    \begin{tikzpicture}[x=0.75pt,y=0.75pt,yscale=-1,xscale=1]
    \draw [color={rgb, 255:red, 0; green, 0; blue, 0 }  ,draw opacity=1 ][line width=1.5]    (146,61.5) -- (178,61.5) -- (254,61.5) ;
    \draw [color={rgb, 255:red, 0; green, 0; blue, 0 }  ,draw opacity=1 ][line width=1.5]    (146,54) -- (146,68.5) ;
    \draw [color={rgb, 255:red, 0; green, 0; blue, 0 }  ,draw opacity=1 ][line width=1.5]    (337,62.5) -- (369,62.5) -- (445,62.5) ;
    \draw [color={rgb, 255:red, 0; green, 0; blue, 0 }  ,draw opacity=1 ][line width=1.5]    (250,163.5) -- (282,163.5) -- (333,163.5) ;
    \draw [color={rgb, 255:red, 0; green, 0; blue, 0 }  ,draw opacity=1 ][line width=1.5]    (421,163.5) -- (453,163.5) -- (504,163.5) ;
    \draw [color={rgb, 255:red, 155; green, 155; blue, 155 }  ,draw opacity=0.47 ][line width=1.5]    (75,163.5) -- (107,163.5) -- (158,163.5) ;
    \draw [color={rgb, 255:red, 0; green, 0; blue, 0 }  ,draw opacity=1 ][line width=1.5]    (180,294.5) -- (212,294.5) -- (448,294.5) ;
    \draw [color={rgb, 255:red, 155; green, 155; blue, 155 }  ,draw opacity=0.55 ][line width=1.5]  [dash pattern={on 5.63pt off 4.5pt}]  (222,38) -- (222,106.5) -- (222,342.5) ;
    \draw [color={rgb, 255:red, 0; green, 0; blue, 0 }  ,draw opacity=1 ][line width=1.5]    (145,89.5) -- (177,89.5) -- (253,89.5) ;
    \draw [color={rgb, 255:red, 0; green, 0; blue, 0 }  ,draw opacity=1 ][line width=1.5]    (145,117.5) -- (177,117.5) -- (253,117.5) ;
    \draw [color={rgb, 255:red, 0; green, 0; blue, 0 }  ,draw opacity=1 ][line width=1.5]    (250,192.5) -- (282,192.5) -- (333,192.5) ;
    \draw [color={rgb, 255:red, 0; green, 0; blue, 0 }  ,draw opacity=1 ][line width=1.5]    (250,222.5) -- (282,222.5) -- (333,222.5) ;
    \draw [color={rgb, 255:red, 0; green, 0; blue, 0 }  ,draw opacity=1 ][line width=1.5]    (145,82) -- (145,96.5) ;
    \draw [color={rgb, 255:red, 0; green, 0; blue, 0 }  ,draw opacity=1 ][line width=1.5]    (145,110) -- (145,124.5) ;
    \draw [color={rgb, 255:red, 0; green, 0; blue, 0 }  ,draw opacity=1 ][line width=1.5]    (337,88.5) -- (369,88.5) -- (445,88.5) ;
    \draw [color={rgb, 255:red, 0; green, 0; blue, 0 }  ,draw opacity=1 ][line width=1.5]    (336,118.5) -- (368,118.5) -- (444,118.5) ;
    \draw [color={rgb, 255:red, 155; green, 155; blue, 155 }  ,draw opacity=0.47 ][line width=1.5]    (75,192.5) -- (107,192.5) -- (158,192.5) ;
    \draw [color={rgb, 255:red, 155; green, 155; blue, 155 }  ,draw opacity=0.47 ][line width=1.5]    (75,221.5) -- (107,221.5) -- (158,221.5) ;
    \draw [color={rgb, 255:red, 0; green, 0; blue, 0 }  ,draw opacity=1 ][line width=1.5]    (421,192.5) -- (453,192.5) -- (504,192.5) ;
    \draw [color={rgb, 255:red, 0; green, 0; blue, 0 }  ,draw opacity=1 ][line width=1.5]    (420,222.5) -- (452,222.5) -- (503,222.5) ;
    \draw [color={rgb, 255:red, 0; green, 0; blue, 0 }  ,draw opacity=1 ][line width=1.5]    (180,262.5) -- (212,262.5) -- (448,262.5) ;
    \draw [color={rgb, 255:red, 0; green, 0; blue, 0 }  ,draw opacity=1 ][line width=1.5]    (180,323.5) -- (212,323.5) -- (448,323.5) ;
    \draw [color={rgb, 255:red, 0; green, 0; blue, 0 }  ,draw opacity=1 ][line width=1.5]    (337,55) -- (337,69.5) ;
    \draw [color={rgb, 255:red, 0; green, 0; blue, 0 }  ,draw opacity=1 ][line width=1.5]    (337,81) -- (337,95.5) ;
    \draw [color={rgb, 255:red, 0; green, 0; blue, 0 }  ,draw opacity=1 ][line width=1.5]    (336,111) -- (336,125.5) ;
    \draw [color={rgb, 255:red, 0; green, 0; blue, 0 }  ,draw opacity=1 ][line width=1.5]    (250,156) -- (250,170.5) ;
    \draw [color={rgb, 255:red, 0; green, 0; blue, 0 }  ,draw opacity=1 ][line width=1.5]    (250,185) -- (250,199.5) ;
    \draw [color={rgb, 255:red, 0; green, 0; blue, 0 }  ,draw opacity=1 ][line width=1.5]    (250,215) -- (250,229.5) ;
    \draw [color={rgb, 255:red, 0; green, 0; blue, 0 }  ,draw opacity=1 ][line width=1.5]    (421,156) -- (421,170.5) ;
    \draw [color={rgb, 255:red, 0; green, 0; blue, 0 }  ,draw opacity=1 ][line width=1.5]    (421,185) -- (421,199.5) ;
    \draw [color={rgb, 255:red, 0; green, 0; blue, 0 }  ,draw opacity=1 ][line width=1.5]    (420,215) -- (420,229.5) ;
    \draw [color={rgb, 255:red, 0; green, 0; blue, 0 }  ,draw opacity=1 ][line width=1.5]    (180,255) -- (180,269.5) ;
    \draw [color={rgb, 255:red, 0; green, 0; blue, 0 }  ,draw opacity=1 ][line width=1.5]    (180,287) -- (180,301.5) ;
    \draw [color={rgb, 255:red, 0; green, 0; blue, 0 }  ,draw opacity=1 ][line width=1.5]    (180,316) -- (180,330.5) ;
    \draw [color={rgb, 255:red, 155; green, 155; blue, 155 }  ,draw opacity=0.47 ][line width=1.5]    (75,156) -- (75,170.5) ;
    \draw [color={rgb, 255:red, 155; green, 155; blue, 155 }  ,draw opacity=0.47 ][line width=1.5]    (75,185) -- (75,199.5) ;
    \draw [color={rgb, 255:red, 155; green, 155; blue, 155 }  ,draw opacity=0.47 ][line width=1.5]    (75,214) -- (75,228.5) ;
    \draw  [line width=1.5]  (247,61.5) .. controls (247,57.63) and (250.13,54.5) .. (254,54.5) .. controls (257.87,54.5) and (261,57.63) .. (261,61.5) .. controls (261,65.37) and (257.87,68.5) .. (254,68.5) .. controls (250.13,68.5) and (247,65.37) .. (247,61.5) -- cycle ; \draw  [line width=1.5]  (247,61.5) -- (261,61.5) ; \draw  [line width=1.5]  (254,54.5) -- (254,68.5) ;
    \draw  [line width=1.5]  (438,62.5) .. controls (438,58.63) and (441.13,55.5) .. (445,55.5) .. controls (448.87,55.5) and (452,58.63) .. (452,62.5) .. controls (452,66.37) and (448.87,69.5) .. (445,69.5) .. controls (441.13,69.5) and (438,66.37) .. (438,62.5) -- cycle ; \draw  [line width=1.5]  (438,62.5) -- (452,62.5) ; \draw  [line width=1.5]  (445,55.5) -- (445,69.5) ;
    \draw  [line width=1.5]  (497,163.5) .. controls (497,159.63) and (500.13,156.5) .. (504,156.5) .. controls (507.87,156.5) and (511,159.63) .. (511,163.5) .. controls (511,167.37) and (507.87,170.5) .. (504,170.5) .. controls (500.13,170.5) and (497,167.37) .. (497,163.5) -- cycle ; \draw  [line width=1.5]  (497,163.5) -- (511,163.5) ; \draw  [line width=1.5]  (504,156.5) -- (504,170.5) ;
    \draw  [line width=1.5]  (326,163.5) .. controls (326,159.63) and (329.13,156.5) .. (333,156.5) .. controls (336.87,156.5) and (340,159.63) .. (340,163.5) .. controls (340,167.37) and (336.87,170.5) .. (333,170.5) .. controls (329.13,170.5) and (326,167.37) .. (326,163.5) -- cycle ; \draw  [line width=1.5]  (326,163.5) -- (340,163.5) ; \draw  [line width=1.5]  (333,156.5) -- (333,170.5) ;
    \draw  [line width=1.5]  (441,262.5) .. controls (441,258.63) and (444.13,255.5) .. (448,255.5) .. controls (451.87,255.5) and (455,258.63) .. (455,262.5) .. controls (455,266.37) and (451.87,269.5) .. (448,269.5) .. controls (444.13,269.5) and (441,266.37) .. (441,262.5) -- cycle ; \draw  [line width=1.5]  (441,262.5) -- (455,262.5) ; \draw  [line width=1.5]  (448,255.5) -- (448,269.5) ;
    \draw  [color={rgb, 255:red, 155; green, 155; blue, 155 }  ,draw opacity=0.47 ][line width=1.5]  (151,163.5) .. controls (151,159.63) and (154.13,156.5) .. (158,156.5) .. controls (161.87,156.5) and (165,159.63) .. (165,163.5) .. controls (165,167.37) and (161.87,170.5) .. (158,170.5) .. controls (154.13,170.5) and (151,167.37) .. (151,163.5) -- cycle ; \draw  [color={rgb, 255:red, 155; green, 155; blue, 155 }  ,draw opacity=0.47 ][line width=1.5]  (151,163.5) -- (165,163.5) ; \draw  [color={rgb, 255:red, 155; green, 155; blue, 155 }  ,draw opacity=0.47 ][line width=1.5]  (158,156.5) -- (158,170.5) ;
    \draw  [line width=1.5]  (253,82) -- (260.5,89.5) -- (253,97) -- (245.5,89.5) -- cycle ;
    \draw  [line width=1.5]  (445,81) -- (452.5,88.5) -- (445,96) -- (437.5,88.5) -- cycle ;
    \draw  [line width=1.5]  (333,185) -- (340.5,192.5) -- (333,200) -- (325.5,192.5) -- cycle ;
    \draw  [line width=1.5]  (504,185) -- (511.5,192.5) -- (504,200) -- (496.5,192.5) -- cycle ;
    \draw  [line width=1.5]  (448,287) -- (455.5,294.5) -- (448,302) -- (440.5,294.5) -- cycle ;
    \draw  [color={rgb, 255:red, 155; green, 155; blue, 155 }  ,draw opacity=0.47 ][line width=1.5]  (158,185) -- (165.5,192.5) -- (158,200) -- (150.5,192.5) -- cycle ;
    \draw  [line width=1.5]  (246,117.5) .. controls (246,113.63) and (249.13,110.5) .. (253,110.5) .. controls (256.87,110.5) and (260,113.63) .. (260,117.5) .. controls (260,121.37) and (256.87,124.5) .. (253,124.5) .. controls (249.13,124.5) and (246,121.37) .. (246,117.5) -- cycle ;
    \draw  [line width=1.5]  (437,118.5) .. controls (437,114.63) and (440.13,111.5) .. (444,111.5) .. controls (447.87,111.5) and (451,114.63) .. (451,118.5) .. controls (451,122.37) and (447.87,125.5) .. (444,125.5) .. controls (440.13,125.5) and (437,122.37) .. (437,118.5) -- cycle ;
    \draw  [line width=1.5]  (326,222.5) .. controls (326,218.63) and (329.13,215.5) .. (333,215.5) .. controls (336.87,215.5) and (340,218.63) .. (340,222.5) .. controls (340,226.37) and (336.87,229.5) .. (333,229.5) .. controls (329.13,229.5) and (326,226.37) .. (326,222.5) -- cycle ;
    \draw  [line width=1.5]  (496,222.5) .. controls (496,218.63) and (499.13,215.5) .. (503,215.5) .. controls (506.87,215.5) and (510,218.63) .. (510,222.5) .. controls (510,226.37) and (506.87,229.5) .. (503,229.5) .. controls (499.13,229.5) and (496,226.37) .. (496,222.5) -- cycle ;
    \draw  [color={rgb, 255:red, 155; green, 155; blue, 155 }  ,draw opacity=0.47 ][line width=1.5]  (151,221.5) .. controls (151,217.63) and (154.13,214.5) .. (158,214.5) .. controls (161.87,214.5) and (165,217.63) .. (165,221.5) .. controls (165,225.37) and (161.87,228.5) .. (158,228.5) .. controls (154.13,228.5) and (151,225.37) .. (151,221.5) -- cycle ;
    \draw  [line width=1.5]  (441,323.5) .. controls (441,319.63) and (444.13,316.5) .. (448,316.5) .. controls (451.87,316.5) and (455,319.63) .. (455,323.5) .. controls (455,327.37) and (451.87,330.5) .. (448,330.5) .. controls (444.13,330.5) and (441,327.37) .. (441,323.5) -- cycle ;
    \draw [line width=1.5]  [dash pattern={on 1.69pt off 2.76pt}]  (253,117.5) -- (287,117.5) ;
    \draw [shift={(290,117.5)}, rotate = 180] [color={rgb, 255:red, 0; green, 0; blue, 0 }  ][line width=1.5]    (14.21,-4.28) .. controls (9.04,-1.82) and (4.3,-0.39) .. (0,0) .. controls (4.3,0.39) and (9.04,1.82) .. (14.21,4.28)   ;
    \draw [line width=1.5]  [dash pattern={on 1.69pt off 2.76pt}]  (444,118.5) -- (478,118.5) ;
    \draw [shift={(481,118.5)}, rotate = 180] [color={rgb, 255:red, 0; green, 0; blue, 0 }  ][line width=1.5]    (14.21,-4.28) .. controls (9.04,-1.82) and (4.3,-0.39) .. (0,0) .. controls (4.3,0.39) and (9.04,1.82) .. (14.21,4.28)   ;
    \draw [line width=1.5]  [dash pattern={on 1.69pt off 2.76pt}]  (333,222.5) -- (367,222.5) ;
    \draw [shift={(370,222.5)}, rotate = 180] [color={rgb, 255:red, 0; green, 0; blue, 0 }  ][line width=1.5]    (14.21,-4.28) .. controls (9.04,-1.82) and (4.3,-0.39) .. (0,0) .. controls (4.3,0.39) and (9.04,1.82) .. (14.21,4.28)   ;
    \draw [line width=1.5]  [dash pattern={on 1.69pt off 2.76pt}]  (503,222.5) -- (537,222.5) ;
    \draw [shift={(540,222.5)}, rotate = 180] [color={rgb, 255:red, 0; green, 0; blue, 0 }  ][line width=1.5]    (14.21,-4.28) .. controls (9.04,-1.82) and (4.3,-0.39) .. (0,0) .. controls (4.3,0.39) and (9.04,1.82) .. (14.21,4.28)   ;
    \draw [line width=1.5]  [dash pattern={on 1.69pt off 2.76pt}]  (448,323.5) -- (482,323.5) ;
    \draw [shift={(485,323.5)}, rotate = 180] [color={rgb, 255:red, 0; green, 0; blue, 0 }  ][line width=1.5]    (14.21,-4.28) .. controls (9.04,-1.82) and (4.3,-0.39) .. (0,0) .. controls (4.3,0.39) and (9.04,1.82) .. (14.21,4.28)   ;
    \draw [color={rgb, 255:red, 155; green, 155; blue, 155 }  ,draw opacity=0.47 ][line width=1.5]  [dash pattern={on 1.69pt off 2.76pt}]  (158,221.5) -- (192,221.5) ;
    \draw [shift={(195,221.5)}, rotate = 180] [color={rgb, 255:red, 155; green, 155; blue, 155 }  ,draw opacity=0.47 ][line width=1.5]    (14.21,-4.28) .. controls (9.04,-1.82) and (4.3,-0.39) .. (0,0) .. controls (4.3,0.39) and (9.04,1.82) .. (14.21,4.28)   ;
    \draw [color={rgb, 255:red, 155; green, 155; blue, 155 }  ,draw opacity=0.55 ][line width=1.5]  [dash pattern={on 5.63pt off 4.5pt}]  (386,38) -- (386,106.5) -- (386,342.5) ;
    
    \draw (217,9.9) node [anchor=north west][inner sep=0.75pt]  [font=\Large,color={rgb, 255:red, 0; green, 0; blue, 0 }  ,opacity=1 ]  {$t$};
    \draw (348,9.9) node [anchor=north west][inner sep=0.75pt]  [font=\Large,color={rgb, 255:red, 0; green, 0; blue, 0 }  ,opacity=1 ]  {$t\ +\ \Delta t$};
    \draw (120,54) node [anchor=north west][inner sep=0.75pt]    {$1a$};
    \draw (120,82) node [anchor=north west][inner sep=0.75pt]    {$1b$};
    \draw (120,110) node [anchor=north west][inner sep=0.75pt]    {$1c$};
    \draw (312,54) node [anchor=north west][inner sep=0.75pt]    {$2a$};
    \draw (312,82) node [anchor=north west][inner sep=0.75pt]    {$2b$};
    \draw (312,110) node [anchor=north west][inner sep=0.75pt]    {$2c$};
    \draw (225,156) node [anchor=north west][inner sep=0.75pt]    {$3a$};
    \draw (225,185) node [anchor=north west][inner sep=0.75pt]    {$3b$};
    \draw (225,215) node [anchor=north west][inner sep=0.75pt]    {$3c$};
    \draw (396,156) node [anchor=north west][inner sep=0.75pt]    {$4a$};
    \draw (396,185) node [anchor=north west][inner sep=0.75pt]    {$4b$};
    \draw (396,215) node [anchor=north west][inner sep=0.75pt]    {$4c$};
    \draw (155,255) node [anchor=north west][inner sep=0.75pt]    {$5a$};
    \draw (155,287) node [anchor=north west][inner sep=0.75pt]    {$5b$};
    \draw (155,316) node [anchor=north west][inner sep=0.75pt]    {$5c$};
    \end{tikzpicture}

    \caption{Relative positions of the ``risk interval" to the interval of interest $[t, t+ \Delta t)$ (indicated by the dashed vertical lines). Risk intervals start at the time of the last negative biopsy (indicated by $\mid$ ) and run until either detection of progression via a positive biopsy ($\oplus$) or the start of early treatment ($\diamondsuit$). In case of right-censoring ($\bigcirc$) the patient remains at risk for either event (indicated by $\rightarrow$). Patients in the grayed out scenarios are excluded from the model evaluation.}
    \label{fig:relpos}
\end{figure}

The patients whose risk interval lies entirely inside the interval of interest and are detected with cancer progression (3a in Figure~\ref{fig:relpos}) are known to be cases (``absolute cases"). Likewise, the patients whose risk interval is located entirely after $t+\Delta t$ (4a, b, and c in Figure~\ref{fig:relpos}) are known to be controls (``absolute controls"). Other types of patients are all potential cases and/or controls. Patients whose follow-up ends before $t$ (shown in grey color in Figure~\ref{fig:relpos}) are excluded from the evaluation.

\subsection{Sensitivity, specificity and AUC} \label{sec:auc}

To calculate the AUC, the ROC curve has to be determined. This can be done by calculating the sensitivities and $1- \text{specificities}$ corresponding to varying values of the discrimination criterion.\citep{Zou2007} Note that in the following, we calculate the metrics to evaluate the predictions for the primary event of interest, cancer progression.

The progression-specific sensitivity for a threshold value $c$, the probability that the predicted risk is larger or equal to $c$ for patients with cancer progression in the interval of interest, $[t, t+\Delta t)$, can be formulated as
\begin{equation}
\begin{aligned}[b]
\text{sen}^{\textsc{prg}}(t, \Delta t, c) &= \Pr\{\Pi^{\textsc{prg}}(t+\Delta t\mid t) \geq c \mid  T^{\textsc{prg*}} \geq t, T^{\textsc{prg*}} <t+\Delta t, T^\textsc{prg*} < T^\textsc{trt*}\} \\
&= \frac{\Pr\{\Pi^{\textsc{prg}}(t+\Delta t\mid t) \geq c, T^{\textsc{prg*}} \geq t, T^{\textsc{prg*}} <t+\Delta t, T^{\textsc{prg*}} < T^{\textsc{trt*}}\}}{\Pr\{T^{\textsc{prg*}} \geq t, T^{\textsc{prg*}} <t+\Delta t, T^{\textsc{prg*}} < T^{\textsc{trt*}}\}}, 
\end{aligned}
\label{eq:sens}
\end{equation}
where $c$ is a value in $[0, 1]$. Similar definitions can be found in \citet{Blanche2013, Blanche2014}. Analogously, following Equation (1) in \citet{Blanche2013}, the specificity, used to evaluate the ability to identify the negative instances, is defined as
\begin{equation}
\begin{aligned}[b]
\text{spe}(t, \Delta t, c) &= \Pr\{\Pi^{\textsc{prg}}(t+\Delta t\mid t)<c \mid \min(T^{\textsc{prg*}}, T^{\textsc{trt*}}) \geq t + \Delta t\} \\
&= \frac{\Pr\{\Pi^{\textsc{prg}}(t+\Delta t\mid t)<c, \min(T^{\textsc{prg*}}, T^{\textsc{trt*}}_i) \geq t + \Delta t\}}{\Pr\{\min(T^{\textsc{prg*}}, T^{\textsc{trt*}}) \geq t + \Delta t\}},
\end{aligned}
\label{eq:sens}
\end{equation}
i.e., the probability that the predicted progression-specific risk for patients who ``survive" the interval of interest is lower than the threshold $c$. In the presence of interval censoring, however, it is often not clear whether a patient is a case, control or even had the event before the interval of interest. We propose two methods to deal with this uncertainty when estimating the progression-specific sensitivity and specificity:
\begin{description}
   \item[Model-based approach] In the model-based approach, all subjects at risk at time $t$ are considered when calculating the sensitivity, but they contribute with different weights, $\mathcal{W}^{\textsc{m}}_i$, that depend on their estimated probability of experiencing cancer progression during the interval of interest for patient $i$ in the test set. The weights are derived from the cumulative incidence function estimated by the model, and are, thus, named model-based weights. The estimated model-based progression-specific sensitivity can be written as
    \begin{align*}
    \widehat{\text{sen}}^{\textsc{prg}}_{\textsc{model}}(t, \Delta t, c) &= \frac{\sum_{i}I\{\Pi^{\textsc{prg}}_i(t+\Delta t\mid t) \geq c\} \times \mathcal{W}^\textsc{m}_i}{\sum_{i}\mathcal{W}^\textsc{m}_i},
    \end{align*}
    where $I(\cdot)$ is the indicator function. For different types of patients, the model-based weights for being a case (visualized in Figure~\ref{fig:senw}; the numbers in parentheses refer to the numbering used in Figure~\ref{fig:relpos}) are specified as
    \begin{align*}
    \mathcal{W}^\textsc{m}_i=
    \begin{cases}
    \frac{\Pi^{\textsc{prg}}_i(T^\textsc{prg+}_i \mid T^{\textsc{prg-}}_i) - \Pi^{\textsc{prg}}_i(t \mid T^{\textsc{prg-}}_i)}{\Pi^{\textsc{prg}}_i(T^\textsc{prg+}_i \mid T^{\textsc{prg-}}_i)} & \text{if }T^{\textsc{prg-}}_i<t, T^{\textsc{prg+}}_i>t, T^{\textsc{prg+}}_i<t+\Delta t \quad \scriptstyle{(1a)}, \\
    \Pi^{\textsc{prg}}_i(T^{\textsc{trt}}_i \mid T^{\textsc{prg-}}_i) - \Pi^{\textsc{prg}}_i(t \mid T^{\textsc{prg-}}_i) & \text{if }T^{\textsc{prg-}}_i<t, T^{\textsc{trt}}_i >t, T^{\textsc{trt}}_i<t+\Delta t \quad \scriptstyle{(1b)},\\
    \Pi^{\textsc{prg}}_i(t+\Delta t \mid T^{\textsc{prg-}}_i)-\Pi^{\textsc{prg}}_i(t \mid T^{\textsc{prg-}}_i) & \text{if }T^{\textsc{prg-}}_i<t, T^{\textsc{cen}}_i>t, T^{\textsc{cen}}_i<t+\Delta t \quad \scriptstyle{(1c)},\\
    \frac{\Pi^{\textsc{prg}}_i(t+\Delta t \mid T^{\textsc{prg-}}_i)}{\Pi^{\textsc{prg}}_i(T^\textsc{prg+}_i \mid T^{\textsc{prg-}}_i)} & \text{if } T^{\textsc{prg-}}_i>t, T^{\textsc{prg-}}_i<t+\Delta t, T^{\textsc{prg+}}_i > t+\Delta t\quad \scriptstyle{(2a)},\\
    \Pi^{\textsc{prg}}_i(t+\Delta t \mid T^{\textsc{prg-}}_i) & \text{if } T^{\textsc{prg-}}_i>t, T^{\textsc{prg-}}_i<t+\Delta t, T^{\textsc{trt}}_i > t+\Delta t\quad \scriptstyle{(2b)},\\
    \Pi^{\textsc{prg}}_i(t+\Delta t \mid T^{\textsc{prg-}}_i) & \text{if } T^{\textsc{prg-}}_i>t, T^{\textsc{prg-}}_i<t+\Delta t, T^{\textsc{cen}}_i > t+\Delta t\quad \scriptstyle{(2c)},\\
    1 & \text{if } T^{\textsc{prg-}}_i>t, T^{\textsc{prg+}}_i<t+\Delta t \quad \scriptstyle{(3a)},\\
    \Pi^{\textsc{prg}}_i(T^{\textsc{trt}}_i \mid T^{\textsc{prg-}}_i) & \text{if } T^{\textsc{prg-}}_i>t, T^{\textsc{trt}}_i<t+\Delta t \quad \scriptstyle{(3b)},\\
    \Pi^{\textsc{prg}}_i(t+\Delta t \mid T^{\textsc{prg-}}_i) & \text{if } T^{\textsc{prg-}}_i>t, T^{\textsc{cen}}_i<t+\Delta t \quad \scriptstyle{(3c)},\\
    \frac{\Pi^{\textsc{prg}}_i(t+\Delta t \mid T^{\textsc{prg-}}_i)-\Pi^{\textsc{prg}}_i(t \mid T^{\textsc{prg-}}_i)}{\Pi^{\textsc{prg}}_i(T^\textsc{prg+}_i \mid T^{\textsc{prg-}}_i)} & \text{if }T^{\textsc{prg-}}_i<t, T^{\textsc{prg+}}_i>t+\Delta t\quad \scriptstyle{(5a)},\\
    \Pi^{\textsc{prg}}_i(t+\Delta t \mid T^{\textsc{prg-}}_i)-\Pi^{\textsc{prg}}_i(t \mid T^{\textsc{prg-}}_i) & \text{if }T^{\textsc{prg-}}_i<t, T^{\textsc{trt}}_i>t+\Delta t\quad \scriptstyle{(5b},\\
    \Pi^{\textsc{prg}}_i(t+\Delta t \mid T^{\textsc{prg-}}_i)-\Pi^{\textsc{prg}}_i(t \mid T^{\textsc{prg-}}_i) & \text{if }T^{\textsc{prg-}}_i<t, T^{\textsc{cen}}_i>t+\Delta t\quad \scriptstyle{(5c)}.\\
    0 & \text{otherwise} \quad \scriptstyle{(4a,4b,4c,\text{ and the grayed-out scenarios})},
    \end{cases}
    \end{align*}
    It is noted that for patient groups 1a, 2a and 5a, since their progression has been detected and must be between the last negative biopsy time $T^\textsc{prg-}_i$ and the positive biopsy time $T^\textsc{prg+}_i$, the weights of being cases are the predicted progression-specific risk differences within the interval of interest rescaled relative to the corresponding risk differences between the last negative biopsy and the positive biopsy. Likewise, the specificity can be estimated correspondingly, by using the model-based weights (visualized in Figure~\ref{fig:spew})
    \begin{align*}
    \widehat{\text{spe}}_{\textsc{model}}(t, \Delta t, c) &= \frac{\sum_{i}I\{\Pi^{\textsc{prg}}_i(t+\Delta t\mid t) < c\} \times \mathcal{W'}^\textsc{m}_i}{\sum_{i}\mathcal{W'}^\textsc{m}_i},
    \end{align*}
    where the model-based weights for defining a control are
    \begin{align*}
    \mathcal{W'}^\textsc{m}_i = 
    \begin{cases}
    \mathcal{S}_i(t+\Delta t \mid T^\textsc{prg-}_i) & \text{if }T^{\textsc{prg-}}_i<t, T^{\textsc{cen}}_i > t, T^{\textsc{cen}}_i<t+\Delta t \quad \scriptstyle{(1c)},\\
    \frac{\Pi^{\textsc{prg}}_i(T^{\textsc{prg+}}_i \mid T^{\textsc{prg-}}_i) - \Pi^{\textsc{prg}}_i(t + \Delta t \mid T^{\textsc{prg-}}_i)}{\Pi^{\textsc{prg}}_i(T^\textsc{prg+}_i \mid T^{\textsc{prg-}}_i)} & \text{if } T^{\textsc{prg-}}_i > t, T^{\textsc{prg-}}_i<t+\Delta t, T^{\textsc{prg+}}_i >t+\Delta t \quad \scriptstyle{(2a)},\\
    1 - \Pi^{\textsc{prg}}_i(t + \Delta t \mid T^{\textsc{prg-}}_i) & \text{if } T^{\textsc{prg-}}_i>t, T^{\textsc{prg-}}_i<t+\Delta t, T^{\textsc{trt}}_i>t+\Delta t \quad \scriptstyle{(2b)},\\
    1 - \Pi^{\textsc{prg}}_i(t + \Delta t \mid T^{\textsc{prg-}}_i) & \text{if } T^{\textsc{prg-}}_i > t, T^{\textsc{prg-}}_i<t+\Delta t, T^{\textsc{cen}}_i > t+\Delta t \quad \scriptstyle{(2c)},\\
    \mathcal{S}_i(t+\Delta t \mid T^\textsc{prg-}_i) & \text{if } T^{\textsc{prg-}}_i > t, T^{\textsc{cen}}_i<t+\Delta t \quad \scriptstyle{(3c)},\\
    1 & \text{if } T^{\textsc{prg-}}_i>t+\Delta t \quad \scriptstyle{(4a, 4b, 4c)}, \\
    \frac{\Pi^{\textsc{prg}}_i(T^{\textsc{prg+}}_i \mid T^{\textsc{prg-}}_i) - \Pi^{\textsc{prg}}_i(t + \Delta t \mid T^{\textsc{prg-}}_i)}{\Pi^{\textsc{prg}}_i(T^\textsc{prg+}_i \mid T^{\textsc{prg-}}_i)} & \text{if } T^{\textsc{prg-}}_i<t, T^{\textsc{prg+}}_i>t+\Delta t \quad \scriptstyle{(5a)},\\
    1 - \Pi^{\textsc{prg}}_i(t + \Delta t \mid T^{\textsc{prg-}}_i) & \text{if } T^{\textsc{prg-}}_i<t, T^{\textsc{trt}}_i>t+\Delta t \quad \scriptstyle{(5b)},\\
    1 - \Pi^{\textsc{prg}}_i(t + \Delta t \mid T^{\textsc{prg-}}_i) & \text{if } T^{\textsc{prg-}}_i<t, T^{\textsc{cen}}_i>t+\Delta t \quad \scriptstyle{(5c)},\\
    0 & \text{otherwise} \quad \scriptstyle{(1a, 1b, 3a, 3b,\text{ and the grayed-out scenarios})}.
    \end{cases}
    \end{align*}
    The model-based weights of being controls for the patients who had progression detected and can potentially be a control (i.e., patient group 2a and 5a) are constructed analogously to their counterpart of being cases: the differences of progression-specific risks between the progression detection time ($T^\textsc{prg+}_i$) and $t+ \Delta t$ scaled relative to the risk differences between the progression detection time ($T^\textsc{prg+}_i$) and the last negative biopsy time ($T^\textsc{prg-}_i$. For patient groups 1c and 3c who were censored before the end of the interval of interest (i.e., $t+\Delta t$), the model-based weights for being controls are the probability of the patient not having either of the events before $t+\Delta t$ conditional on the fact that neither event happened until the last negative biopsy, i.e., the conditional overall survival function $\mathcal{S}_i(t+\Delta t \mid T^\textsc{prg-}_i)$. For patients who were censored or treated after $t+\Delta t$ (i.e., patient groups 2b, 2c, 5b and 5c), it is known that they did not initiate treatment before $t+\Delta t$. Therefore, their weight of being a control is calculated as one minus the conditional progression-specific risk at $t+\Delta t$.
\begin{figure}[H]
\centering
\begin{subfigure}[t]{0.49\textwidth}
    \centering
    \includegraphics[scale=0.22]{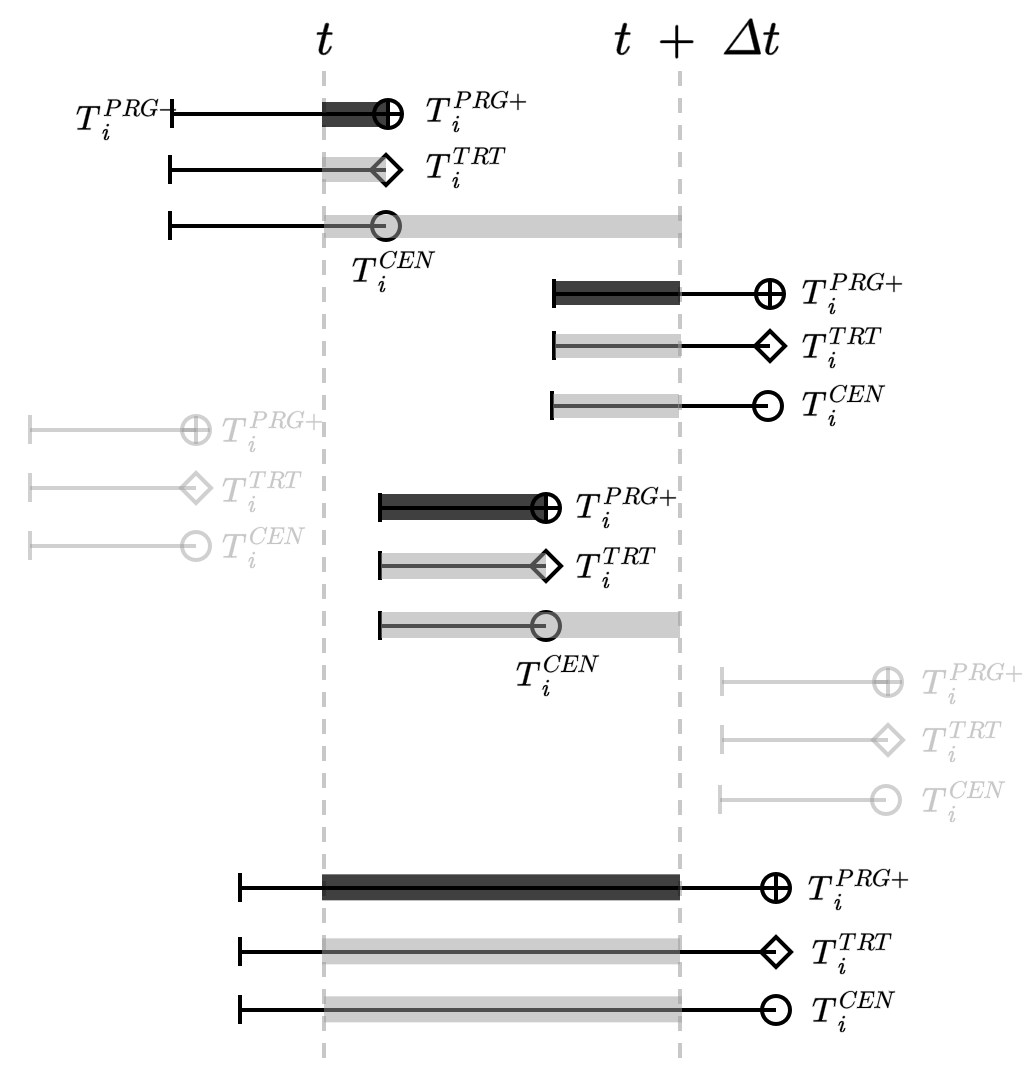}
    \caption{Sensitivity}
    \label{fig:senw}
\end{subfigure}
\begin{subfigure}[t]{0.49\textwidth}
    \centering
    \includegraphics[scale=0.22]{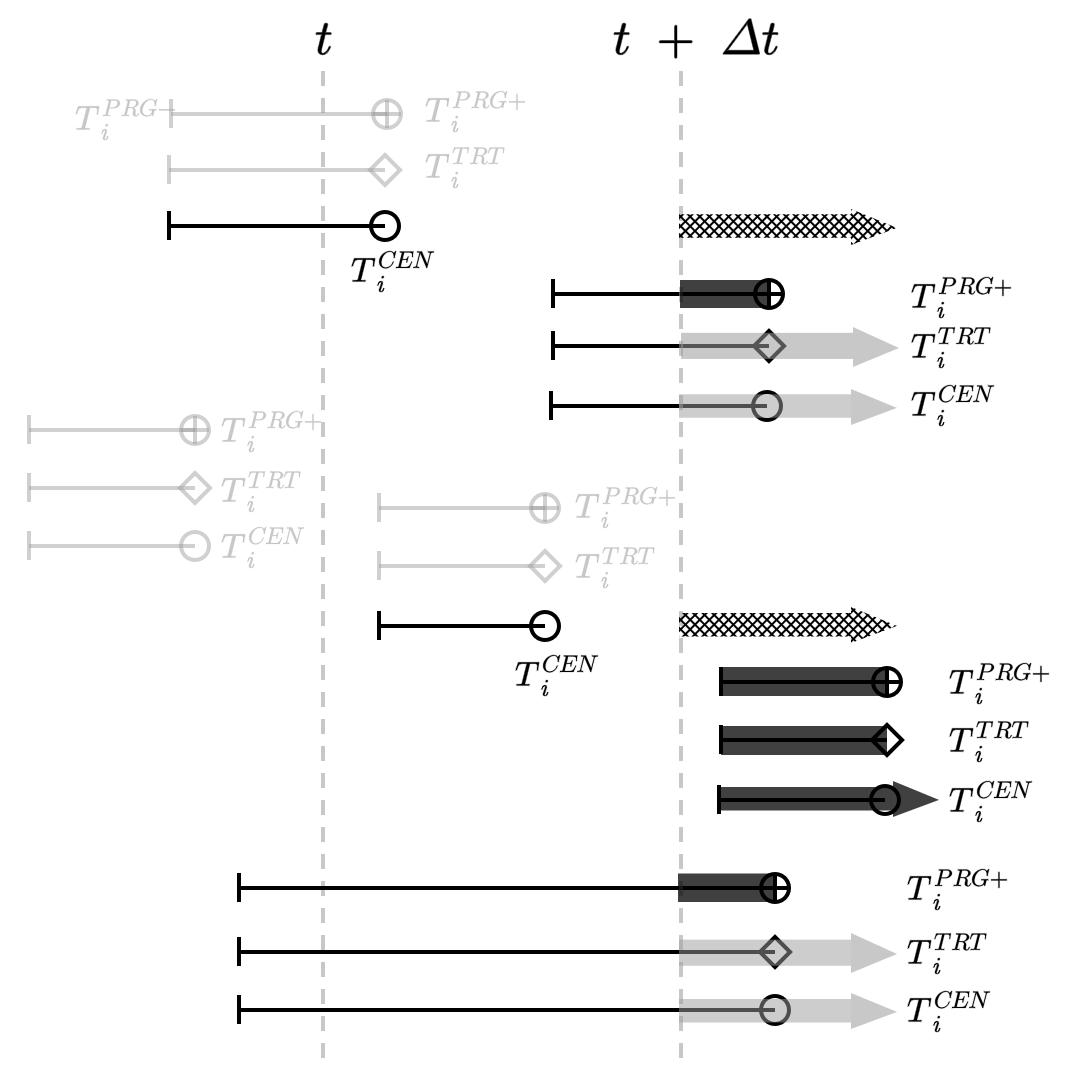}
    \caption{Specificity}
    \label{fig:spew}
\end{subfigure}
\caption{Illustration of the model-based weights for sensitivity (a) and specificity (b) in calculating AUC and Brier score. The shaded bars represent the time with which the patients contribute to being a case (a) or a control (b). The lightgray shades are the weights using the predicted progression-specific risks. The dark shades indicate using the rescaled progression-specific risks. The crosshatches indicate using the overall survival function. Grayed-out patient groups are not considered in the calculation.}
\end{figure}
    
    \item[IPCW approach] Another way to calculate the progression-specific sensitivity is to utilize only the subset of patients for whom the event is known to be in the interval of interest, i.e., the absolute cases (3a in Figure~\ref{fig:relpos}), and weigh them to also represent the patients who were censored before experiencing the primary event of interest. These weights are the inverse of the probability of not being censored before time t, obtained using the Kaplan-Meier (KM) estimator. The estimated IPCW-based progression-specific sensitivity can be written as 
    \begin{align*}
    \widehat{\text{sen}}^{\textsc{prg}}_{\textsc{ipcw}}(t, \Delta t, c) &= \frac{\sum_{i:T^{\textsc{prg-}}_i \geq t , T^{\textsc{prg+}}_i<t+\Delta t}I\{\Pi^{\textsc{prg}}_i(t+\Delta t\mid t) \geq c\} \times \mathcal{W}^\textsc{ipcw}_i}{\sum_{i:T^{\textsc{prg-}}_i \geq t , T^{\textsc{prg+}}_i<t+\Delta t}\mathcal{W}^\textsc{ipcw}_i},
    \end{align*}
    where $\mathcal{W}^\textsc{ipcw}_i = \{G(T^{\textsc{prg+}}_i \mid t)\}^{-1}$, and $G(s \mid t)$ is the conditional probability of a patient being censoring-free at $s$ given being uncensored at $t$. Furthermore, to estimate the IPCW-based specificity, the subset of the patients known to be event-free at $t+\Delta t$, i.e., the absolute controls (4a, 4b and 4c in Figure~\ref{fig:relpos}), is used
    \begin{align}
    \widehat{\text{spe}}_{\textsc{ipcw}}(t, \Delta t, c) &= \frac{\sum_{i:T^{\textsc{prg-}}_i>t+\Delta t}I\{\Pi^{\textsc{prg}}_i(t+\Delta t\mid t)  <c\}  \times \mathcal{W'}^\textsc{ipcw}_i}{\sum_{i:T^{\textsc{prg-}}_i>t+\Delta t}\mathcal{W'}^\textsc{ipcw}_i}.
    \label{eqn:ipcwspe}
    \end{align}
    The corresponding weights are estimated using the KM estimator for being censoring-free at time $t + \Delta t$, to represent the patients censored during the interval of interest. Thus the weights are the inverse of the conditional probability of being censoring-free at the end of the interval of interest, $\mathcal{W'}^\textsc{ipcw}_i = \{G(t+\Delta t \mid t)\}^{-1}$. As these weights are uniform for all the controls and will unsurprisingly cancel out, (\ref{eqn:ipcwspe}) can be simplified to 
    \begin{align*}
    \widehat{\text{spe}}_{\textsc{ipcw}}(t, \Delta t, c) &= \frac{\sum_{i:T^{\textsc{prg-}}_i>t+\Delta t}I\{\Pi^{\textsc{prg}}_i(t+\Delta t\mid t)  <c\}}{\sum_{i:T^{\textsc{prg-}}_i>t+\Delta t}1}.
    \end{align*}
\end{description}

Using either of the versions of sensitivity and specificity, we can calculate the AUC within $[t, t+\Delta t)$, $\text{AUC}(t, \Delta t)$, as
\begin{align*}
    \text{AUC}(t, \Delta t)=\int^1_0\text{sen}^{\textsc{prg}}(t, \Delta t, c)d\{1-\text{spe}^{\textsc{prg}}(t, \Delta t, c)\}.
\end{align*}

\subsection{Brier score}

The Brier score is an evaluation metric that combines discrimination and calibration. \citep{Gerds2006,Steyerberg2009} It quantifies how close the predicted probabilities are to the actual binary outcomes, with a lower score indicating better model performance. The progression-specific Brier score has the following formula
\begin{align*}
\text{BS}^\textsc{prg}(t+\Delta t, t) = E\left[\left\{I(T^\textsc{prg*}_i < t + \Delta t) - \Pi^\textsc{prg}_i(t+\Delta t \mid t )\right\}^2  \mid T^{(2)}_i \geq t \right],
\end{align*}
where $T^{(2)}_i$ is $T^{\textsc{cen}}_i$, $T^{\textsc{prg+}}_i$ or $T^{\textsc{trt}}_i$ whichever was observed for patient $i$.
The patient who is a case (i.e., experiencing cancer progression during the interval of interest) is desired to have a probability of predicted risk closer to one, whereas for a control the predicted risk of progression should be low. The Brier score of the test set can be estimated using the model-based approach as well as the IPCW approach, using the same subsets of patients and weights as for the AUC.
\begin{description}
   \item[Model-based approach] Utilizing the set of patients at risk of cancer progression at time $t$ and the model-based weights given in Section~\ref{sec:auc}, the Brier score can be estimated by 
    \begin{align*}
    \widehat{\text{BS}}^\textsc{prg}_{\textsc{model}}(t+\Delta t, t) = \frac{1}{n_t}\sum_{i: T^{(2)}_i \geq t}\left[\left\{1 - \Pi^\textsc{prg}_i(t+\Delta t \mid t )\right\}^2 \times \mathcal{W}^\textsc{m}_i + \left\{0 - \Pi^\textsc{prg}_i(t+\Delta t \mid t )\right\}^2 \times \mathcal{W'}^\textsc{m}_i\right],
    \end{align*}
    where $n_t$ is the number of patients who have not 
    been detected with progression, been treated, or censored until time $t$. 
    
    \item[IPCW approach] Analogously, using the IPC weights (as specified in Section~\ref{sec:auc}), we have 
    \begin{align*}
    \widehat{\text{BS}}^\textsc{prg}_{\textsc{IPCW}}(t+\Delta t, t) = \frac{1}{n_t} &\Bigg[\sum_{i:T^{\textsc{prg-}}_i \geq t, T^{\textsc{prg+}}_i<t+\Delta t}\left\{1 - \Pi^\textsc{prg}_i(t+\Delta t \mid t )\right\}^2 \times \mathcal{W}^\textsc{ipcw}_i + \\
    &\ \ \ \sum_{i:T^{\textsc{prg-}}_i>t+\Delta t}\left\{0 - \Pi^\textsc{prg}_i(t+\Delta t \mid t )\right\}^2 \times \mathcal{W'}^\textsc{ipcw}_i\Bigg].
    \end{align*}
    For calculating the first factor, only the subjects detected with cancer progression within the interval of interest are used whereas for the second factor, only the subjects who are event-free until $t+\Delta t$ contribute via their weights.
\end{description}

\subsection{Expected predictive cross-entropy (EPCE)}

The EPCE is an evaluation quantity from information theory, which describes the expected value of the cross-entropy between the true and predicted risk distributions. \citep{Commenges2012} In our setting, we focus on the density of cancer progression risk. The progression-specific EPCE has the following form
\begin{align*}
\text{EPCE}^{\textsc{prg}}(t+\Delta t, t) = E\left\{-\log\left[p\left\{T^{\textsc{prg}*}_i \mid T^{\textsc{prg}*}_i \geq t, T^{\textsc{prg}*}_i < t+\Delta t, T^{\textsc{prg}*}_i < T^{\textsc{trt}*}_i, \boldsymbol{\mathcal{Y}}_i(t), \boldsymbol{\mathcal{D}}_n\right\}\right]\right\}.
\end{align*}
The EPCE can be estimated by
\begin{align*}
\widehat{\text{EPCE}}^{\textsc{prg}}(t+\Delta t, t) = \frac{1}{n_t}\sum_{i: T^{(2)}_i \geq t}-\log\left[p\{\tilde{T}^{(1)}_i, \tilde{T}^{(2)}_i, \tilde{\delta}^{(1)}_i, \tilde{\delta}^{(2)}_i \mid T^{\textsc{prg}*}_i \geq t, T^{\textsc{prg}*}_i < T^{\textsc{trt}*}_i, \boldsymbol{\mathcal{Y}}_i(t), \boldsymbol{\mathcal{D}}_n\}\right],
\end{align*}
where $\tilde{T}^{(1)}_i = \max(T^\textsc{prg-}_i, t)$, 
\begin{align*}
\tilde{T}^{(2)}_i = 
\begin{cases} 
    \min(T^\textsc{prg+}_i, t+\Delta t), &\delta_i = 1,  \\
    \min(T^\textsc{trt}_i, t+\Delta t), &\delta_i = 2, \\
    t+\Delta t, &\delta_i = 0,
\end{cases}
\end{align*}
$\tilde{\delta}^{(1)}_i = I(T^{\textsc{prg-}}_i \leq t +\Delta t, T^{(2)}_i \geq t)$, and $\tilde{\delta}^{(2)}_i = I\left(T^{\textsc{prg-}}_i \geq t +\Delta t, \delta_i=0\right)$. The term in the log function can be estimated by
\begin{align*}
p\{\tilde{T}^{(1)}_i, \tilde{T}^{(2)}_i, \tilde{\delta}^{(1)}_i, \tilde{\delta}^{(2)}_i \mid T^{\textsc{prg}*}_i \geq t, T^{\textsc{prg}*}_i < T^{\textsc{trt}*}_i, \boldsymbol{\mathcal{Y}}_i(t), \boldsymbol{\mathcal{D}}_n\} = \log\left[\tilde{\delta}^{(1)}_i \mathcal{F}_1 + \tilde{\delta}^{(2)}_i \mathcal{F}_2 \right],
\end{align*}
where the first factor, $\mathcal{F}_1 = \displaystyle{\int}^{\tilde{T}^{(2)}_i}_{\tilde{T}^{(1)}_i}\Pr\{\tilde{T}^{(1)}_i \leq T^{\textsc{prg}*}_i < s \mid T^{\textsc{prg}*}_i \geq t, T^{\textsc{prg}*}_i < T^{\textsc{trt}*}_i, \boldsymbol{\mathcal{Y}}_i(t), \boldsymbol{\mathcal{D}}_n\}ds$, is the cumulative incidence function of progression over $\tilde{T}^{(1)}_i$ and $\tilde{T}^{(2)}_i$ , and the second factor, $\mathcal{F}_2 = \frac{\Pr\{T^*_i \geq t+\Delta t \mid \boldsymbol{\mathcal{Y}}_i(t), \boldsymbol{\mathcal{D}}_n\}}{\Pr\{T^{*}_i \geq t \mid \boldsymbol{\mathcal{Y}}_i(t), \boldsymbol{\mathcal{D}}_n\}}$, is the overall survival probability of the patient being event-free at $t + \Delta t$ conditional on that he did not experience either of the events until $t$.

\section{Application to the ICJM} \label{sec:app}


To illustrate the practical application of these methods, we analyze a joint model that can handle interval censoring and competing risks fitted on data from the Canary Prostate Active Surveillance Study. This study focuses on the progression of prostate cancer to a stage necessitating intervention, with progression detected through interval-censored biopsies. Additionally, the initiation of early treatment, which occurs before the cancer reaches a critical stage, is considered a competing risk. The patients were examined with biopsies according to the PASS schedules (at 12, 24 months, and biennially afterwards) to detect the primary event of interest, cancer progression, resulting in interval censoring. Around 10\% of the patients left AS for treatment before cancer progression was detected, constituting the competing event.\citep{Newcomb2016} In addition, the repeatedly measured biomarker, PSA, is a time-varying predictor for the events of interest. We developed the ICJM 1 in \citet{Yang2023} from the Canary PASS data (more details can be found in Web Appendix 2). The repeatedly measured PSA levels, for patient $j$ in the training set, were modelled in the longitudinal submodel of the ICJM, a mixed-effects model,
\begin{align*}
\log_2\{\mbox{PSA}_j(t) + 1\} = \beta_0 + u_{0j} + \displaystyle{\sum}^3_{p=1}(\beta_p + u_{pj})\mathcal{C}^{(p)}_j(t) + \beta_4(\mbox{Age}_j - 62) + \epsilon_j(t), \quad \boldsymbol{u}_j \sim N(\mathbf{0}, \boldsymbol{\Omega}), 
\end{align*}
where $\mathcal{C}(\cdot)$ is the design matrix for the natural cubic splines, $\mbox{Age}_j$ refers to the patient's age at the start of active surveillance, and $\boldsymbol{\beta} = (\beta_0, \dots, \beta_4)^\top$ is a vector of corresponding regression coefficients. The error terms $\epsilon_j(t)$ were assumed to follow a Student's t-distribution with three degrees of freedom, and the random effects, $\boldsymbol{u}_j = \{u_{0i}, u_{1i}, u_{2i}, u_{3i}\}^\top$, were assumed to be normally distributed with a mean of zero and an unstructured variance-covariance matrix $\boldsymbol{\Omega}$. The expected values of $\log_2\{\mbox{PSA}_j(t) + 1\}$, which we denote by $m_j(t)$, were then incorporated into a survival submodel using two functional forms:
\begin{align*}
h_j^{(k)}\left\{t \mid \boldsymbol{\mathcal{M}}_j(t), \mbox{PSAD}_j\right\} = h_0^{(k)}(t)\exp\left[\gamma_k^\top\mbox{PSAD}_j + \alpha_{1k} m_j(t) + \alpha_{2k}\{m_j(t) - m_j(t - 1)\}\right],
\end{align*}
where the hazard of the $j$-th patient in the training set for event $k$ ($k \in \boldsymbol{\mathcal{K}}=\{\textsc{prg}, \textsc{trt}\}$, where $\textsc{prg}$ stands for cancer progression and $\textsc{trt}$ for early treatment) at time $t$ is denoted as $h_j^{(k)}(t)$. The vector $\boldsymbol{\mathcal{M}}_j(t) = \{\boldsymbol{m}_j(s); 0 \leq s < t\}$, contains the estimated PSA trajectories until $t$, with the corresponding regression coefficient $\alpha_k$ for the functional forms of estimated PSA levels, and $\mbox{PSAD}_j$ is the baseline PSA density (calculated as the baseline PSA level divided by the baseline prostate gland's volume), with corresponding regression coefficient $\gamma_k$. The model included two functional forms of the estimated PSA levels, the estimated PSA level at $t$, $m_j(t)$ and the yearly change in the estimated PSA, $m_j(t) - m_j(t - 1)$. The event $k$-specific baseline hazard $h_0^{(k)}(t)$ was specified with the penalized B-splines:
\begin{align*}
    \log h_0^{(k)}(t) = \gamma_{k, h_0, 0} + \sum^A_{a=1}{\gamma_{k, h_0, a}\mathcal{G}_a(t, \boldsymbol{\xi})},
\end{align*}
where $\mathcal{G}_a(t, \boldsymbol{\xi})$ is the $a$-th basis function of a B-splines with knots $\xi_1, \dots, \xi_A$. The number of knots was chosen to be 11, and the penalized coefficients for the basis function are denoted by $\boldsymbol{\gamma}_{k, h_0}$. The ICJM was estimated in the Bayesian framework. We evaluated the predictive performance of the ICJM for the interval of interest, $(t, t + \Delta t] = (1,4]$, i.e., using covariate information up until year 1. The evaluation was performed internally on the Canary PASS data using 5-fold cross-validation. A total of 812 (out of 833) subjects at risk in year one were included in the evaluation. During the interval of interest, 50 subjects started the competing event, early treatment. The model-based approach resulted in an AUC of 0.64 while the AUC estimated using the IPCW approach was higher, namely 0.68. Also in the calculation of the Brier score, the model-based approach estimated a less favorable performance than the IPCW approach, namely 0.17 versus 0.07. The model's estimated EPCE was -0.72.



\section{Simulation} \label{sec:sim}

\subsection{Simulation setting}
As the results of the application in Section~\ref{sec:app} show, the model-based and IPCW approach can result in relatively different estimates of AUC and Brier score. To investigate the performance of the two methods in a more controlled setting, we conducted two simulation studies. In both studies, the accuracy metrics were calculated using the longitudinal information (i.e., PSA) until year one in the test set and evaluating the performance up until year four, i.e., the interval of interest being $[1, 4)$. 200 datasets were simulated based on the ICJM handling interval censoring of biopsies and competing risk of early treatment described in Section~\ref{sec:app} and the PASS biopsy schedule \citep{Newcomb2016}, each with 300 subjects. In the 200 datasets, there are on average 22\% patients who were observed with cancer progression, 9\% who initiated early treatment.

\subsection{Comparison between model specification}

Since the model-based approach re-uses the prediction model for its evaluation, model misspecification might lead to overestimation of the model's predictive performance. Therefore, in this simulation, we aim to explore the influence of model misspecification on the accuracy metrics from both approaches. Three models, including the correctly-specified model and two misspecified models (i.e., ignoring one baseline covariate in the survival submodel and wrongly assuming a linear evolution for the PSA trajectory), were fitted on each dataset (as the training set). Each fitted model was then evaluated on another simulated dataset (as the test set in which we utilize the PSA information until year one for predictions). As a reference, we also calculated AUC, Brier score, and EPCE, based on the true event times, i.e., for the ideal scenario where there is no censoring, and it is, thus, known whether a patient is a case or a control. It is noted that the reference metrics were calculated using the same risk estimates from the corresponding model when compared to the model-based and IPCW approach. Furthermore, we also presented a naive calculation of AUC and Brier score, i.e., ignoring the interval censoring essence of cancer progression and using the observed progression time to determine a case or control.

\begin{figure}[H]
\centering
\begin{subfigure}{0.32\textwidth}
    \centering
    \includegraphics[width=\linewidth]{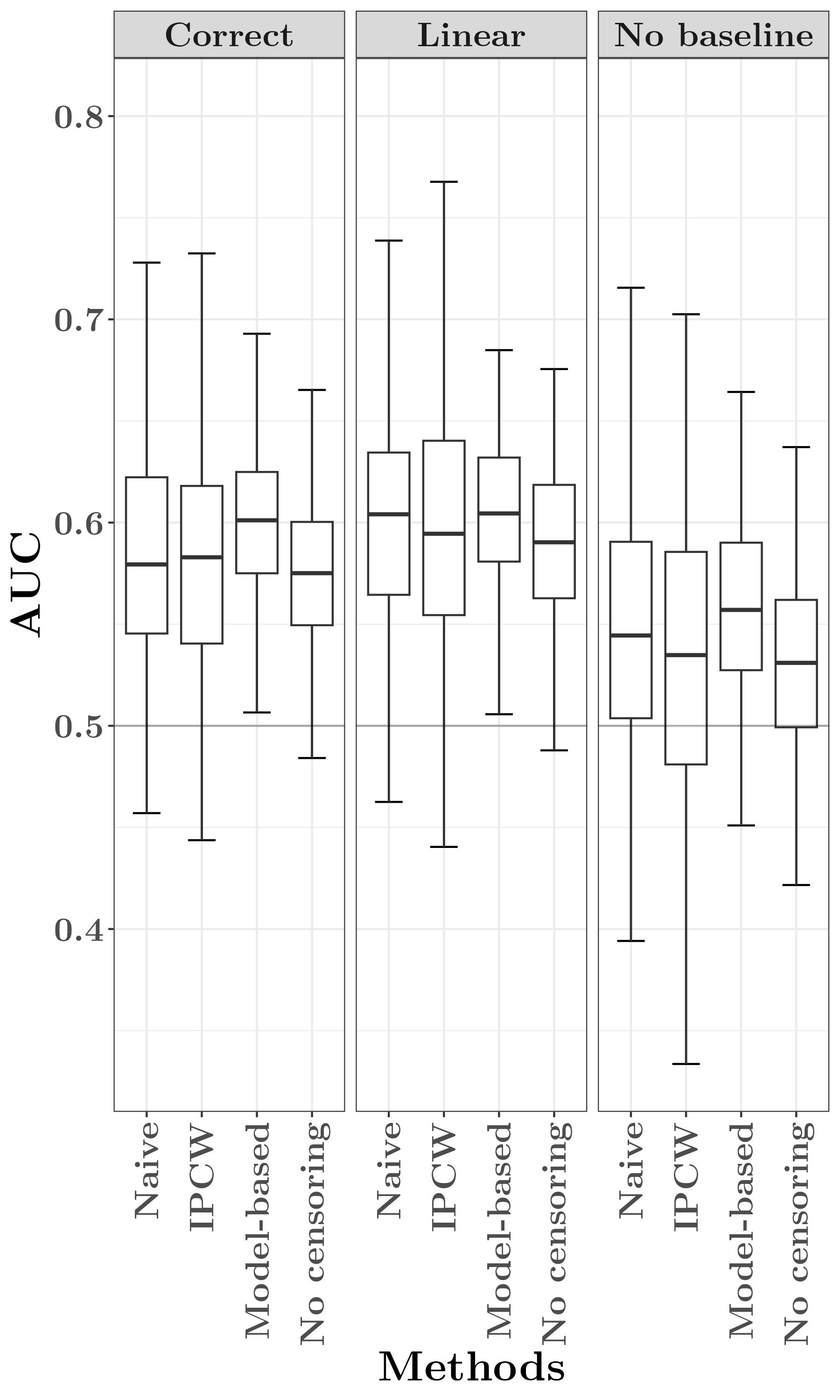}
    \caption{AUC}
    \label{fig:aucmodel}
\end{subfigure}
\begin{subfigure}{0.32\textwidth}
    \centering
    \includegraphics[width=\linewidth]{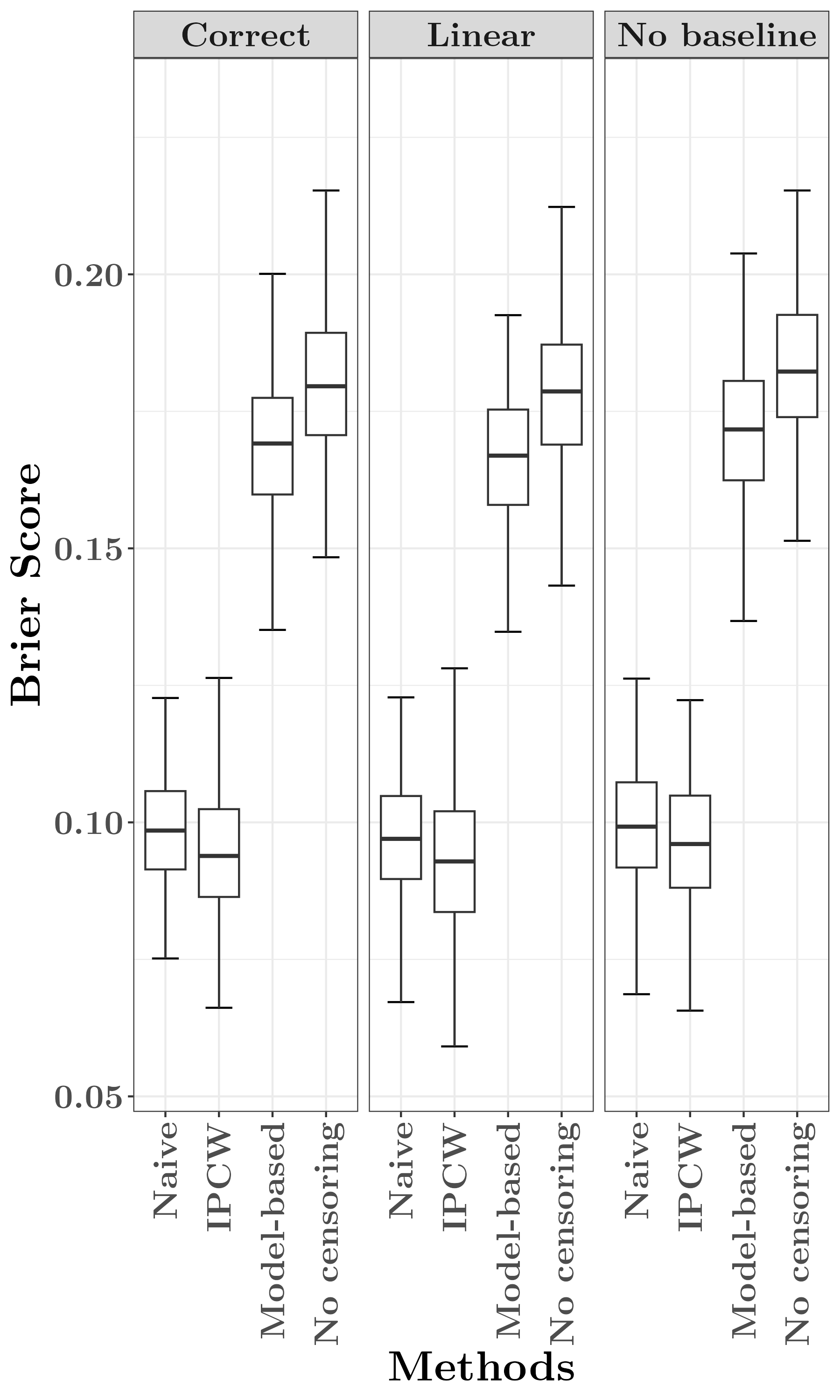}
    \caption{Brier score}
    \label{fig:bsmodel}
\end{subfigure}
\begin{subfigure}{0.32\textwidth}
    \centering
    \includegraphics[width=\linewidth]{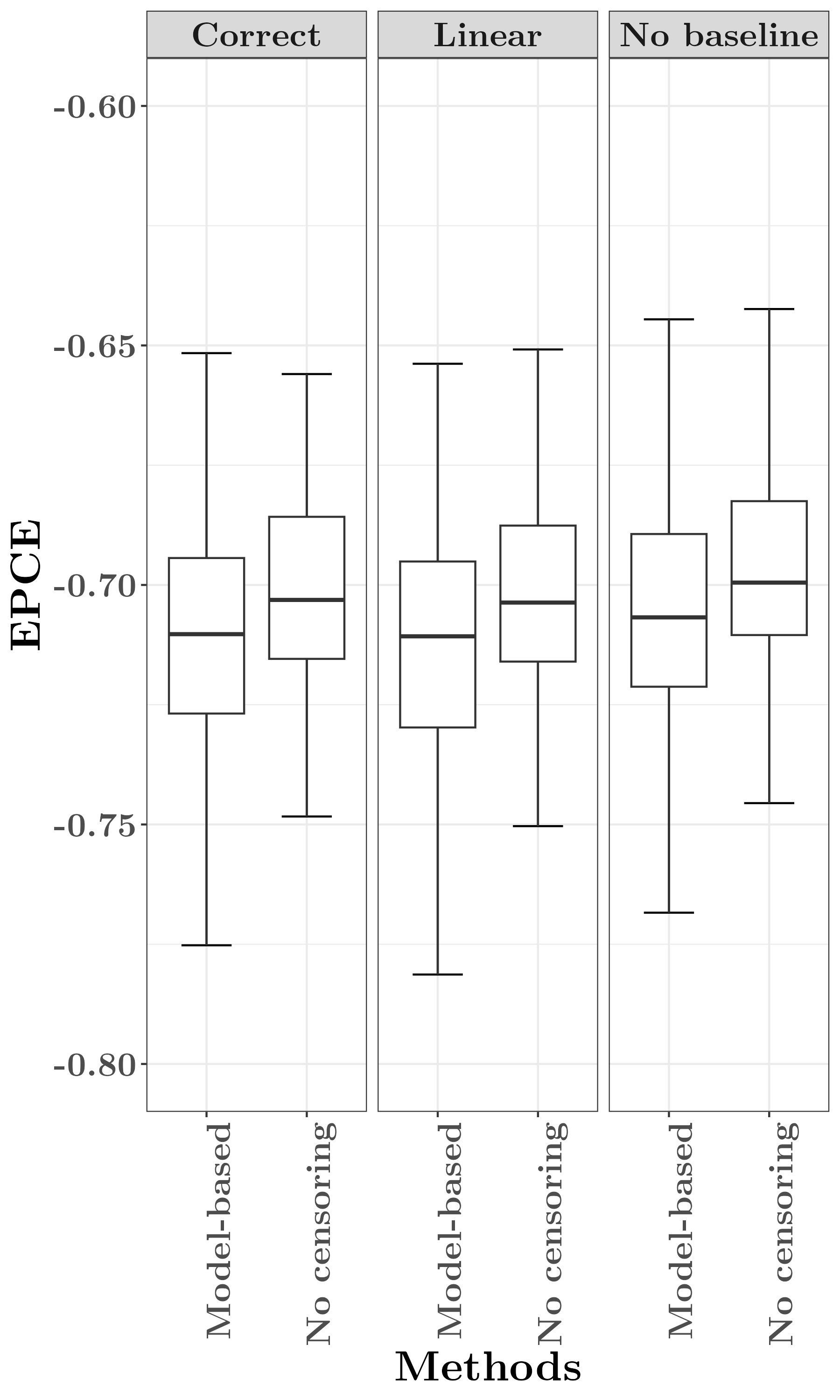}
    \caption{EPCE}
    \label{fig:epcemodel}
\end{subfigure}
\caption{The model-based, IPCW and naive (ignoring interval censoring) estimates of the AUC (a) and Brier score (b), and the model-based estimates of the EPCE (c) from the correctly-specified model, the model with a linear PSA trajectory, and the model ignoring the baseline covariate (PSA density), compared to the corresponding ideal scenario when there is no censoring.}
\label{fig:model}
\end{figure}

The main results from the simulation study are shown in Figure~\ref{fig:model} and Table~\ref{tab:model}. Comparison of the AUC between the three models for the ideal scenario without censoring (Figure~\ref{fig:aucmodel}) shows that misspecification of the effect of time on PSA as linear had no relevant effect on the model's performance whereas omitting the baseline covariate worsened the discriminative ability considerably. Simply ignoring interval censoring resulted in, on average, relatively small differences in the AUC compared to the ideal scenario in which the exact event times are known. However, the variability in this approach was larger than in the model-based approach. The naive approach performed worse than the model-based approach but was comparable to the IPCW approach in estimating the Brier score for all three models whether misspecified or not. The IPCW estimates of the AUC generally had greater variability whereas the median estimates across all three models were closer to the reference AUC than the median of the model-based estimates. The model-based approach tended to overestimate the AUC. Misspecification of the model by omitting the baseline covariate resulted in a larger RMSE for the model-based approach. However, the variability of the model-based AUC across simulations was not inflated. Analogous to the AUC, the model-based Brier scores for all three models had less variability but were, on average, closer to the reference Brier score than the one from the IPCW approach. Only the model-based approach generated estimates of the Brier score that were similar to the ideal scenario for all three models. The model-based approach also had the lowest RMSE in estimating both the AUC and Brier scores for all three models, approximately 38\% better for the AUC and 83\% better for the Brier score compared to the IPCW approach, and approximately 24\% better for the AUC and 82\% better for the Brier score compared to the naive approach. The model-based estimates of the EPCE for the three models were comparable to the reference EPCE (see Figure~\ref{fig:epcemodel}). The reference EPCE also did not vary much among different models.

\begin{table}[H]
\resizebox{0.4\linewidth}{!}{
    \begin{subtable}[c]{0.48\textwidth} 
        \centering
        \subcaption{AUC}
        \label{tab:aucmodel}
        \begin{tabular}{cccc}
          \toprule
          & Model-based & IPCW & Naive  \\
          \midrule
          Correctly-specified & 0.036 & 0.056 & 0.045\\
          Linear & 0.029 & 0.054 & 0.043 \\
          No baseline covariate & 0.040 & 0.058 & 0.051\\
          \bottomrule
        \end{tabular}
    \end{subtable}
}
    \hspace{0.1\textwidth} 
\resizebox{0.4\linewidth}{!}{
    \begin{subtable}[c]{0.48\textwidth} 
        \centering
        \subcaption{Brier score}
        \label{tab:bsmodel}
        \begin{tabular}{cccc}
          \toprule
          & Model-based & IPCW & Naive  \\
          \midrule
          Correctly-specified & 0.015 & 0.087 & 0.083 \\
          Linear & 0.015 & 0.086 & 0.082 \\
          No baseline covariate & 0.015 & 0.088 & 0.085 \\
          \bottomrule
        \end{tabular}
    \end{subtable}
}
    \caption{The root mean square error (RMSE) of the model-based, IPCW and naive (ignoring interval censoring) estimates of the AUC (a) and Brier score (b) from the correctly-specified model, the model with a PSA trajectory, and the model ignoring the baseline covariate (PSA density), to the corresponding reference metrics without censoring.}
    \label{tab:model}
\end{table}

\subsection{Comparison between biopsy frequencies}

\begin{figure}[H]
\centering
\includegraphics[scale=0.35]{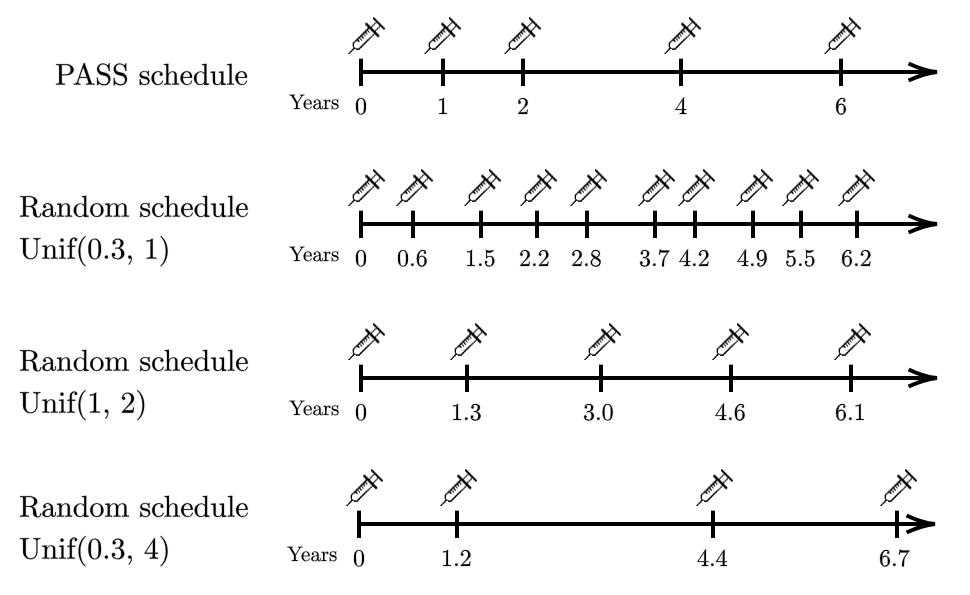}
\caption{The visualizations of one example series of schedules in the second simulation.}
\label{fig:schdvis}
\end{figure}

The uncertainty due to interval censoring is closely related to the frequency of examination, in our case, the biopsies. With more recurrent biopsies, the true event times are restricted to shorter intervals and, thus, are expected to increase the predictive performance. Therefore, this second simulation study aims to explore the potential effect of the information intensity from the interval censoring on different approaches of estimating the AUC and Brier score. Again, we simulated the datasets based on the ICJM 1 but applied four different biopsy schedules, namely the PASS schedule, and three schedules with random biopsy intervals, uniformly distributed between 0.3 and 1 year, 1 and 2 years, and 0.3 and 4 years, respectively (see examples for different scenarios in Figure~\ref{fig:schdvis}). For simplicity, we only compare the estimates of the AUC and Brier score from the correctly-specified model in this case.

\begin{figure}[H]
\centering
\begin{subfigure}[t]{0.49\textwidth}
    \centering
    \includegraphics[scale=0.28]{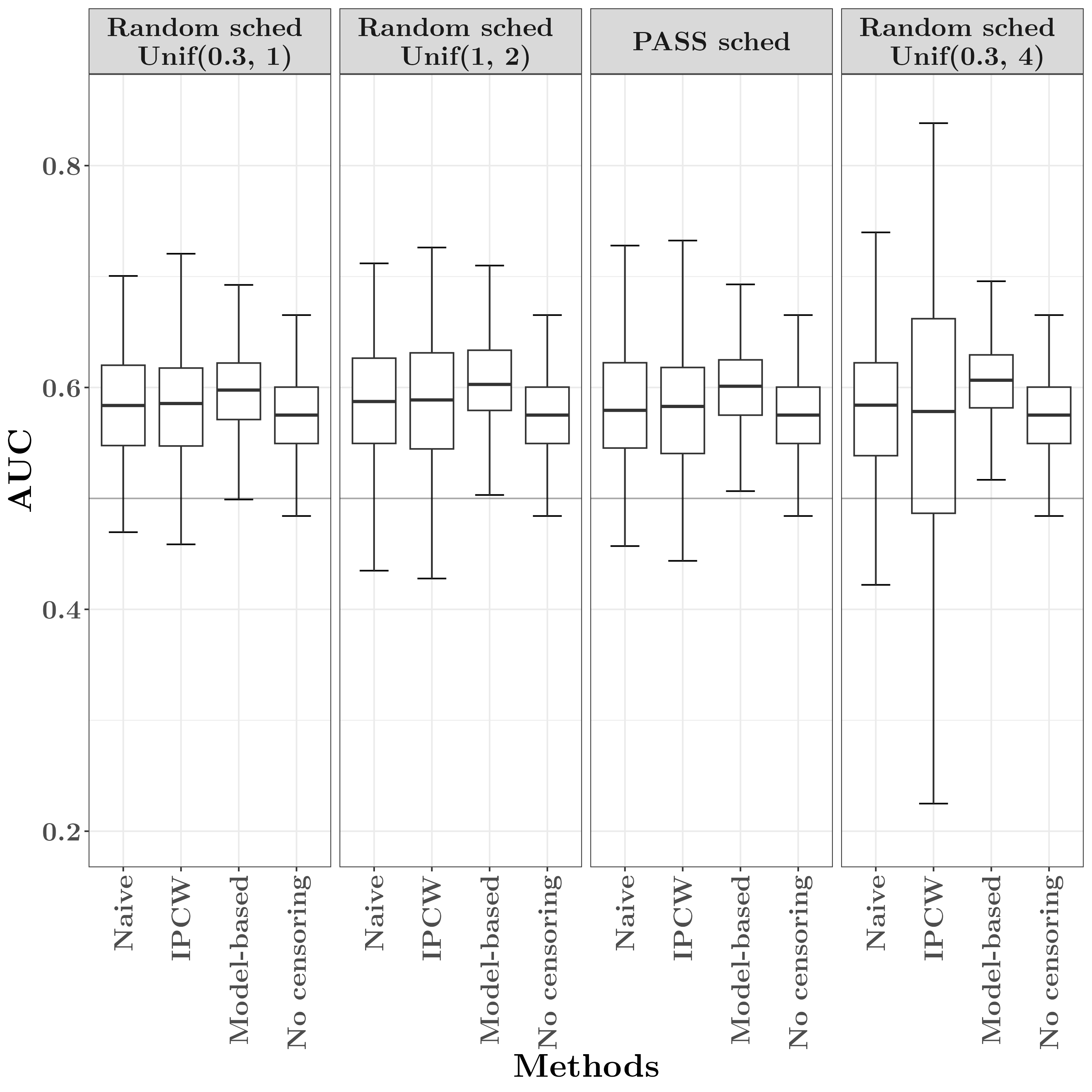}
    \caption{AUC}
    \label{fig:aucschd}
\end{subfigure}
\begin{subfigure}[t]{0.49\textwidth}
    \centering
    \includegraphics[scale=0.28]{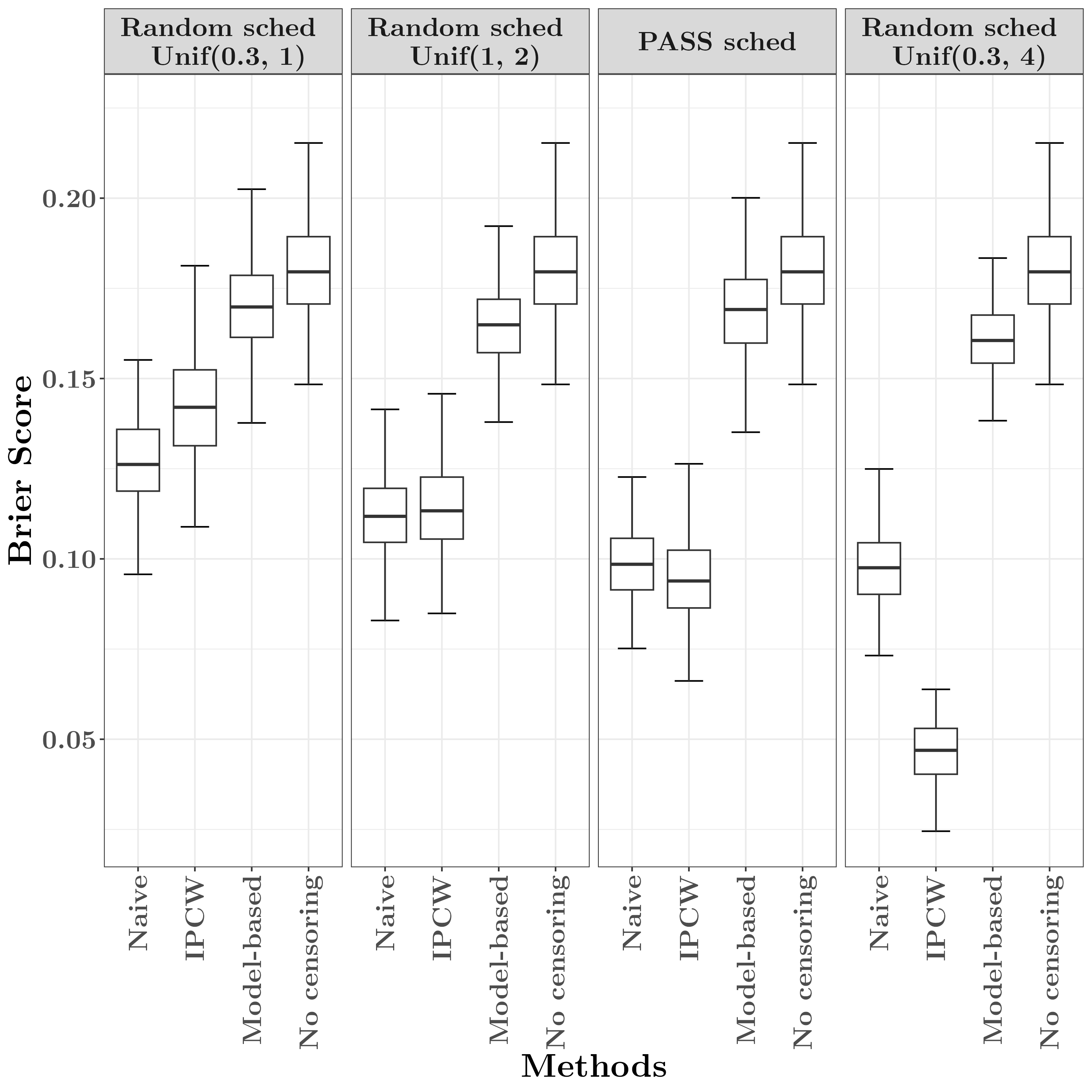}
    \caption{Brier score}
    \label{fig:bsschd}
\end{subfigure}
\caption{The model-based, IPCW and naive (ignoring interval censoring) estimates of the AUC (a) and Brier score (b) from the correctly-specified model, in the context of a PASS biopsy schedule, a random schedule from 0.3 to 1 year, a random schedule from 1 to 2 years, and a random schedule from 0.3 to 4 years, compared to the corresponding ideal scenario when there is no censoring.}
\label{fig:schd}
\end{figure}

The estimated AUC and Brier scores from three approaches and different biopsy frequencies are depicted in Figure~\ref{fig:schd}, and the corresponding calculated RMSE are listed in Table~\ref{tab:schd}. With shorter biopsy intervals (i.e., more frequent biopsies), the IPCW estimates of the AUC tended to have lower variability, whereas the variability in the model-based approach was less impacted by the biopsy schedules. This inflation of the variability with less frequent biopsies was observed in the naive approach as well, but was less obvious than in the IPCW approach. With larger biopsy intervals, i.e., fewer biopsies, the bias in the model-based AUC estimates increased while the median of the IPCW estimates was always closer to that of the ideal AUC and remained more stable across different biopsy schedules. All three approaches had similar RMSE of the AUC estimates in the most frequent biopsy schedule. The model-based approach, on average for the four schedules, improved the RMSE of AUC estimates by 38\% relative to the IPCW approach and 10\% relative to the naive approach. The model-based Brier scores were much closer to those from the ideal scenario without censoring and were more robust to lower biopsy frequencies. The Brier score estimates from the IPCW and naive approach differed a lot from those from the ideal scenario, especially with larger biopsy intervals. This tendency was much more manifest in the IPCW Brier scores. Regarding the RMSE of the Brier score, the model-based approach was on average 79\% better than the IPCW approach and 76\% better than the naive approach. The variability of all Brier score estimates was robust to the biopsy schedules. 

\begin{table}[H]
\resizebox{0.37\linewidth}{!}{
    \begin{subtable}[c]{0.48\textwidth} 
        \centering
        \subcaption{AUC}
        \label{tab:aucschd}
        \begin{tabular}{cccc}
          \toprule
          & Model-based & IPCW & Naive  \\
          \midrule
          Random sched Unif(0.3,1) & 0.038 & 0.040 & 0.039 \\
          Random sched Unif(1,2) & 0.045 & 0.054 & 0.046\\
          PASS sched & 0.036 & 0.056 & 0.045  \\
          Random sched Unif(0.3,4) & 0.047 & 0.118 & 0.054 \\
          \bottomrule
        \end{tabular}
    \end{subtable}
}
    \hspace{0.13\textwidth} 
\resizebox{0.37\linewidth}{!}{
    \begin{subtable}[c]{0.48\textwidth} 
        \centering
        \subcaption{Brier score}
        \label{tab:bsschd}
        \begin{tabular}{cccc}
          \toprule
          & Model-based & IPCW & Naive \\
          \midrule
          Random sched Unif(0.3,1) & 0.013 & 0.039 & 0.055 \\
          Random sched Unif(1,2) & 0.019 & 0.067 & 0.070 \\
          PASS sched & 0.015 & 0.087 & 0.083  \\
          Random sched Unif(0.3,4) & 0.023 & 0.135 & 0.084 \\
          \bottomrule
        \end{tabular}
    \end{subtable}
}
    \caption{The root mean square error (RMSE) of the model-based, IPCW and naive (interval censoring) estimates of the AUC (a) and Brier score (b) from the correctly-specified model, in the context of a PASS biopsy schedule, a random schedule from 0.3 to 1 year, a random schedule from 1 to 2 years, and a random schedule from 0.3 to 4 years, to the corresponding reference metrics without censoring.}
    \label{tab:schd}
\end{table}

\section{Discussion} \label{sec:disc}

In this study, we presented two approaches, to estimate commonly used time-dependent predictive accuracy metrics, the AUC, Brier score and EPCE, for a prediction model or algorithm in a setting with competing risks and interval censoring: a model-based and an IPCW approach. The two approaches follow different strategies to deal with the uncertainty about a patient's true event status at a particular time of interest due to censoring. The model-based approach uses all subjects in the risk set for evaluation and weighs their contributions to the estimation based on the estimated probability of experiencing the event (i.e., being a case) or ``surviving" the interval of interest (i.e., being a control). The IPCW approach only utilizes those subjects in the estimation that are known to be a case or control. Our simulation study showed that the IPCW estimates typically have greater variability but are more robust to model misspecification. The model-based approach was more sensitive to the model misspecification but more robust to the frequency of the examinations (resulting in interval censoring). A possible reason is that for sparser biopsy schedules, not only the uncertainty about the true event time is larger, but, as a result, fewer patients will be included in the IPCW approach since it is less likely that the risk interval between a negative and positive biopsy is fully contained in the interval of interest. In most of the scenarios, especially the Brier score and the combination of precision and variability (i.e., RMSE), the model-based approach outperformed the IPCW approach. The EPCE can only be calculated for the model-based approach. Even with interval censoring and competing risks, the model-based EPCE can recover the reference EPCE (i.e., in the ideal scenario without any censoring).

One notable point when using time-dependent accuracy metrics is the choice of the interval of interest. For the IPCW approach, it is necessary to have both cases (i.e., the interval between a negative and a positive examination, must be fully contained in the interval of interest) and controls (i.e., the negative examination is after the end of the interval of interest). Therefore, the length of the interval of interest has to be longer than the minimal length of the examination interval. In the case of a fixed schedule like the Canary PASS, the timing of the biopsies should also be taken into account. Furthermore, the model is expected to have better predictive performance for later intervals of interest since more longitudinal information is used in that case. On the other hand, when the interval of interest is late in the follow-up, there will be few(er) ``absolute controls". In our first simulation study, it was noticed that the linear model was highly comparable to the correctly-specified model. This is because the predictions were based on the PSA measurements in the first year, and the simulated (overall nonlinear) PSA trajectories were relatively linear during the first year (see Figure S1).

As with all statistical methods, the two approaches presented here have strengths and limitations. The model-based approach typically has less variability and performed much better than the IPCW approach in estimating the Brier score. In addition, the model-based approach is capable of handling informative censoring that depends on the history of the longitudinal outcome(s), which the IPCW approach does not directly accommodate.\citep{Rizopoulos2024} However, it tends to be overly optimistic since the estimated risk probabilities from the model itself are used in its evaluation, making this approach's performance more dependent on the correct model specification. On the other hand, the IPCW-based AUC estimates were less biased, but this approach has the disadvantage of a potentially large information loss, particularly in the setting with interval censoring where there may be only a few ``absolute cases" and/or ``absolute controls". This issue is exacerbated when the examinations are infrequent and/or the interval of interest is too long. The IPCW approach requires an additional assumption that the timing of the examination (in our case, the biopsies) are not dependent on any covariate information or previously observed repeated measurements. This assumption is not necessary for the model-based approach. In our application, the assumption holds, as the biopsy schedule was predefined in the Canary PASS protocol. Other than the non-parametric way to calculate IPC weights, using semiparametric models including important biomarker information as covariates is an alternative to calculate the patient-specific IPC weights.\citep{Blanche2013}

Our evaluation of the investigated methods has some limitations. The simulation study was set up to evaluate the performance of the two methods in our own real-world application. In that setting, our choices were restricted by the fact that most of the detected progression occurred around year three and that biopsies around the most cancer progression were performed in years one, two and four. The interval of interest should start earlier than year two or end later than year four to ensure the possibility of having ``absolute cases". It is expected to improve the predictive performance if the interval of interest could be chosen to contain enough ``absolute cases" and ``absolute controls", and simultaneously take advantage of enough longitudinal information. However, we added the second simulation to explore how these results can be generalized to other settings in which the intervals between examinations may be different. Moreover, in this study, we assumed the sensitivity of biopsies to be perfect, meaning that the underlying cancer progression can only occur between a negative and a positive biopsy. Nevertheless in real practice, there exist many periodic examinations with low sensitivity, indicating that the underlying event(s) might have occurred even before the negative examination. This brings extra challenge to the underlying timing of interval censored event(s), and consequently leads to difficulty in defining cases and controls. Our previous study solved this problem in the model estimation phase.\citep{Yang2024} Similarly, the corresponding weights in the accuracy metrics could be calculated as a weighted sum of the event-specific risks in each related examination period multiplied by the misclassification probability determined by a pre-specified test sensitivity.

To conclude, the model-based approach is a reliable method in terms of bias and variability for evaluating prediction models with time-varying covariates, competing risks, and interval censoring, especially when interval censoring results in losing the information of a subject being a case or control that is essential in the IPCW approach.

\section*{Acknowledgement}
The research was funded by the National Institutes of Health (the NIH CISNET Prostate Award CA253910). The authors would also like to show our gratitude to the Canary PASS team and all study participants.

\section*{Conflict of Interest}

The authors have declared no conflict of interest.

\bibliographystyle{abbrvnat}
\bibliography{ref}

\end{document}


\maketitle

\section{Canary PASS Data}

\subsection{Rationale of the competing risk, early treatment}

AS is acknowledged as a way to reduce overtreatment for prostate cancer. Low-risk cancer patients are admitted to AS where they are monitored by regular biopsies and (more frequent) blood testing. The (often invasive) treatment, such as radiation or prostatectomy, is deferred until cancer progression (to a Gleason score $\geq7$) is detected. However, not everyone strictly follows this protocol. Some patients initiate treatment earlier, without cancer progression being detected, commonly due to personal reasons, such as anxiety of carrying cancer that might progress, change in preference between regular biopsies or the curative treatment.\citep{Tosoian2016, Beckmann2021} This early treatment is considered as the competing event in this study as cancer progression can no longer be observed once early treatment is initiated.

\section{The Interval-censored Cause-specific Joint model (ICJM)}

The ICJM mentioned in the manuscript handles the competing risk from early treatment and interval censoring of cancer progression due to periodic biopsies via the likelihood. The likelihood of the survival part for the patient $j$ in the training set can be written as:
\begin{align*}
    p(\boldsymbol{T}_j, \delta_j \mid \boldsymbol{u}_j, \boldsymbol{\theta}) &= \left[\exp\left\{-
    \int^{T_j^{\textsc{prg-}}}_0 h^{(\textsc{prg})}_j(\nu)d\nu - \int^{T_j^{\textsc{cen}}}_0 h^{(\textsc{trt})}_j(\nu)d\nu\right\}\right]^{I(\delta_j = 0)} \\  
    &\ \  \times \Bigg[\int^{T^{\textsc{prg+}}_j}_{T^{\textsc{prg-}}_j}h^{(\textsc{prg})}_j(s) \exp\Bigg\{- \int^s_0h^{(\textsc{prg})}_j(\nu)d\nu \\
    &\qquad\qquad\qquad\qquad\qquad - \int^{T^{\textsc{prg+}}_j}_0 h^{(\textsc{trt})}_j(\nu)d\nu\Bigg\}ds\Bigg]^{I(\delta_j = 1)} \\
    &\ \  \times \Bigg[h^{(\textsc{trt})}_j(T^{\textsc{trt}}_j) \exp\Bigg\{-\int^{T_j^{\textsc{prg-}}}_0h^{(\textsc{prg})}_j(\nu)d\nu \\
    &\qquad\qquad\qquad\qquad\qquad -\int^{T^{\textsc{trt}}_j}_0 h^{(\textsc{trt})}_j(\nu)d\nu\Bigg\}\Bigg]^{I(\delta_j = 2)},
\end{align*}
where $\boldsymbol{T}_j$ is the time vector for patient $j$ that may include the last negative biopsy time $T^{\textsc{prg-}}_j$, and censoring time $T^{\textsc{cen}}_j$, first positive biopsy time (cancer progression detection time) $T^{\textsc{prg+}}_j$, or treatment initiation time $T^{\textsc{trt}}_j$; $\delta_j=\{0,1,2\}$ is the observed event indicator with $0$ for the censored patients, $1$ for the patients detected with cancer progression and $2$ for the patients who started early treatment; $h^{(\textsc{prg})}_j(\cdot)$ and $h^{(\textsc{prg})}_j(\cdot)$ are the progression- and treatment-specific instantaneous hazards. The first factor (for $\delta_j = 0$), is the probability of not having experienced any event up until the respective event-free times. Since it is only known that cancer progression did not happen until the last biopsy, patients contribute to the ``overall survival" part of the likelihood only until their event-specific event-free times, the last biopsy time $T^{\textsc{prg-}}_j$ and censoring time ($T^{\textsc{cen}}_j$). The second factor (for $\delta_j = 1$) models the probability of patients having progression in the interval between the last progression-free biopsy, $T^{\textsc{prg-}}_j$, and the biopsy at which progression was detected, $T^{\textsc{prg+}}_j$. For those patients the ``overall survival" part includes the probability that patients are progression-free until time $s$, where $s$ ranges over the interval $(T^{\textsc{prg-}}_j, T^{\textsc{prg+}}_j]$, as well as the probability that the patient did not initiate early treatment before the detection of progression, i.e., until $T^{\textsc{prg+}}_j$. The third factor (for $\delta_j = 2$) uses the standard cause-specific cumulative incidence function in the ``overall survival" part to formulate the probability of patients starting early treatment at $T^{\textsc{trt+}}_j$, conditional on patients not having progressed until the last biopsy (i.e., progression-free time) $T^{\textsc{prg-}}_j$. More details can be found in \cite{Yang2023}.

\subsection{Prediction}

Using the baseline covariates and time-varying biomarker information up until time $t$, we can calculate the probability of the new patient $i$ (in the test dataset) experience the event of interest (in this case, cancer progression) before $t+\Delta t$ conditional on that they do not have either event until $t$, $\Pi^\textsc{prg}_i(t+\Delta t \mid t)$:
\begin{align*}
    \Pi^{\textsc{prg}}_{i}(t+\Delta t\mid t) &= \Pr\Bigg[T^{\textsc{prg}*}_{i} \leq t + \Delta t, T^{\textsc{prg}*}_{i} < T^{\textsc{trt}*}_{i} \mid T^{\textsc{prg}*}_{i} > t, T^{\textsc{trt}*}_{i} > t, \boldsymbol{\mathcal{X}}_{i}(t), \boldsymbol{\mathcal{D}}_n\Bigg] \\
    &= \int\int \Pr\Bigg\{T^{\textsc{prg}*}_{i} \leq t + \Delta t, T^{\textsc{prg}*}_{i} < T^{\textsc{trt}*}_{i} \mid T^{\textsc{prg}*}_{i} > t, T^{\textsc{trt}*}_{i} > t, \boldsymbol{u}_{i}, \boldsymbol{\theta} \Bigg\} \\
    &\qquad\qquad p\left\{\boldsymbol{u}_{i} \mid T^{\textsc{prg}*}_{i} > t, T^{\textsc{trt}*}_{i} > t, \boldsymbol{\mathcal{X}}_{i}(t), \boldsymbol{\theta}\right\} \\
    &\qquad\qquad p(\boldsymbol{\theta} \mid \boldsymbol{\mathcal{D}}_n) d\boldsymbol{u}_{i} d\boldsymbol{\theta},
\end{align*}
where the first term inside the integral can be rewritten based on the Bayes rule as
\begin{align*}
    &\Pr\Bigg\{T^{\textsc{prg}*}_{i} \leq t + \Delta t, T^{\textsc{prg}*}_{i} < T^{\textsc{trt}*}_{i} \mid T^{\textsc{prg}*}_{i} > t, T^{\textsc{trt}*}_{i} > t, \boldsymbol{u}_{i}, \boldsymbol{\theta} \Bigg\} \\
    &=\frac{\Pr\left\{t<T^{\textsc{prg}*}_{i} \leq t+\Delta t, \max(T^{\textsc{prg}*}_{i},t) < T^{\textsc{trt}*}_{i} \mid \boldsymbol{u}_{i}, \boldsymbol{\theta}\right\}}{\Pr\left\{T^{\textsc{prg}*}_{i} > t, T^{\textsc{trt}*}_{i} > t\mid \boldsymbol{u}_{i}, \boldsymbol{\theta}\right\}} \\
    &=\frac{\displaystyle{\int}^{t+\Delta t}_{t} h^{(\textsc{prg})}_{i}(\nu)\exp\left[-\int^{\nu}_0 h^{(\textsc{prg})}_{i}(\nu)-\int^{\nu}_0 h^{(\textsc{trt})}_{i}\{\nu\}\right]d\nu}{\exp\left[-\int^t_0 h^{(\textsc{prg})}_{i}\{t\}-\int^t_0 h^{(\textsc{trt})}_{i}(t)\right]},
\end{align*}
The nested integrals in the above equation do not have a closed-form solution, and can be numerically approximated using the 15-point Gauss-Kronrod rule. Inference can be performed using the following Monto Carlo sampling scheme.\citep{Rizopoulos2011_dympred}

\section{Simulation}

\subsection{Simulation setting}

We simulated data based on the parameters from the ICJM (see Section 3 of the manuscript) fitted on the Canary PASS data. The patients in this study were scheduled to undergo biopsies in months 12, 24 and afterwards biennially and to have PSA measurements taken every three months. The ICJM had the following structure:
\begin{align*}
    \log_2(\text{PSA}_j + 1)(t) &= m_{\textsc{psa},j}(t) + \epsilon_j(t),\\
                    m_{\textsc{psa},j}(t)  &= \beta_0 + u_{0j} + \sum^3_{p=1}(\beta_p + u_{pj})\mathcal{C}^{(p)}_j(t) + \beta_4(\text{Age}_j - 62), \\
    h_j^{(k)}\left\{t \mid \boldsymbol{\mathcal{M}}_{\textsc{psa},j}(t)\right\} &= h_0^{(k)}(t)\exp\Big[\gamma_k\text{density}_j + f\left\{\boldsymbol{\mathcal{M}}_{\textsc{psa},j}(t), \boldsymbol{\alpha}_{k}\right\}\Big],
\end{align*}
where $\mathcal{C}(t)$is the design matrix for the natural cubic splines (with three degrees of freedom) for time $t$; $\text{Age}_j$ and $\text{density}_j$ refer to the patient's age and PSA density at the start of active surveillance. Baseline age was centered by subtracting the median age (62 years) for computational reasons. Both the expected value of PSA and the change in expected PSA over the previous year (where extrapolation was conducted for time points earlier than year one) were included as covariates in the time-to-event component, i.e.,
\begin{align*}
    f\left\{\boldsymbol{\mathcal{M}}_{\textsc{psa},j}(t), \boldsymbol{\alpha}_{k}\right\} = &\  \alpha_{1k,\textsc{psa}}m_{\textsc{psa},j}(t) + \alpha_{2k,\textsc{psa}}\Big\{m_{\textsc{psa},j}(t) - m_{\textsc{psa},j}(t-1)\Big\}.
\end{align*}

The residuals of the longitudinal component were assumed to follow a Student's t distribution with three degrees of freedom \citep{Ani2022},
\begin{align*}
    \epsilon_j(t) \sim t(\frac{1}{\tau_\epsilon}, 3),
\end{align*}
with
\begin{align*}
    \tau_\epsilon \sim \text{Gamma}(0.01, 0.01).
\end{align*}
The prior distributions for the regression coefficients were specified as vague normal distributions,
\begin{align*}
    \beta &\sim \mathcal{N}(0, 100), \\
    \gamma_k &\sim \mathcal{N}(0, 100), \\
    \alpha_{1k,\textsc{psa}}, \alpha_{2k,\textsc{psa}} &\sim \mathcal{N}(0, 100),
\end{align*}
and the variance-covariance matrix of the random effects, $\boldsymbol{\Omega}$, to follow an inverse-Wishart distribution,
\begin{align*}
    \boldsymbol{\Omega} \sim \mathcal{IW}(n_u + 1, \frac{4}{\tau_u}),
\end{align*}
with
\begin{align*}
    \tau_u \sim \text{Gamma}(0.5, 0.01),
\end{align*}
where $n_u$ is the number of parameters in the variance-covariance matrix of the random effects.

The model was implemented in JAGS \citep{JAGS} and run for 10000 iterations, using a thinning interval of 10, in each of three MCMC chains.

The resulting posterior means used for simulation were
\begin{align*}
    \boldsymbol{\beta} &= [2.34, 0.28, 0.61, 0.95, 0.02]^\top, \\
    \boldsymbol{\Omega} &= \begin{bmatrix}
        0.48 & -0.04 & -0.07 & 0.02 \\
        -0.04 & 0.77 & 0.46 & -0.04 \\
        -0.07 & 0.46 & 1.37 & 1.36 \\
        0.02 & -0.04 & 1.36 & 2.54
    \end{bmatrix}, \\
    \tau_\epsilon &= 47.40, \\
    \boldsymbol{\gamma}_{h_0} &= \begin{bmatrix}
        -6.78 & -5.76 \\
        -4.72 & -4.99 \\
        -2.84 & -4.43 \\
        -1.65 & -4.26 \\
        -1.54 & -4.36 \\
        -1.79 & -4.47 \\
        -1.85 & -4.60 \\
        -1.75 & -4.69 \\
        -1.85 & -4.78 \\
        -2.04 & -4.92 \\
        -2.18 & -5.08 \\
        -2.32 & -5.21
    \end{bmatrix},\\
    \boldsymbol{\gamma} &= [0.50, 0.23], \\
    \boldsymbol{\alpha} &= \begin{bmatrix} 
        0.13 & 0.42 \\
        3.01 & 2.62
    \end{bmatrix}.
\end{align*}
The resulting simulated data matched the observed data well with regard to the rates of cancer progression, early treatment initiation and censoring (Table~\ref{Tab:summarytraining}).

\begin{table}[H]
    \centering
    \caption{Summary of event proportions in the simulated training datasets compared to the observed data.}
    \begin{tabular}{lcc}
        \toprule
        \textbf{Events}  & \textbf{Simulated data}$^\dagger$ (\%) & \textbf{Observed data} (\%) \\ \midrule
        Cancer progression & 22.35 & 21.97 \\
        Treatment & 9.18 & 10.44 \\
        Censoring & 68.47 & 67.59 \\
        \bottomrule
        \multicolumn{3}{l}{$^\dagger$: the average proportions overall 200 datasets are presented.}
    \end{tabular}
    \label{Tab:summarytraining}
\end{table}

\subsection{Misspecification of the linear model}

\begin{figure}[H]
\centering
\includegraphics[width=0.9\linewidth]{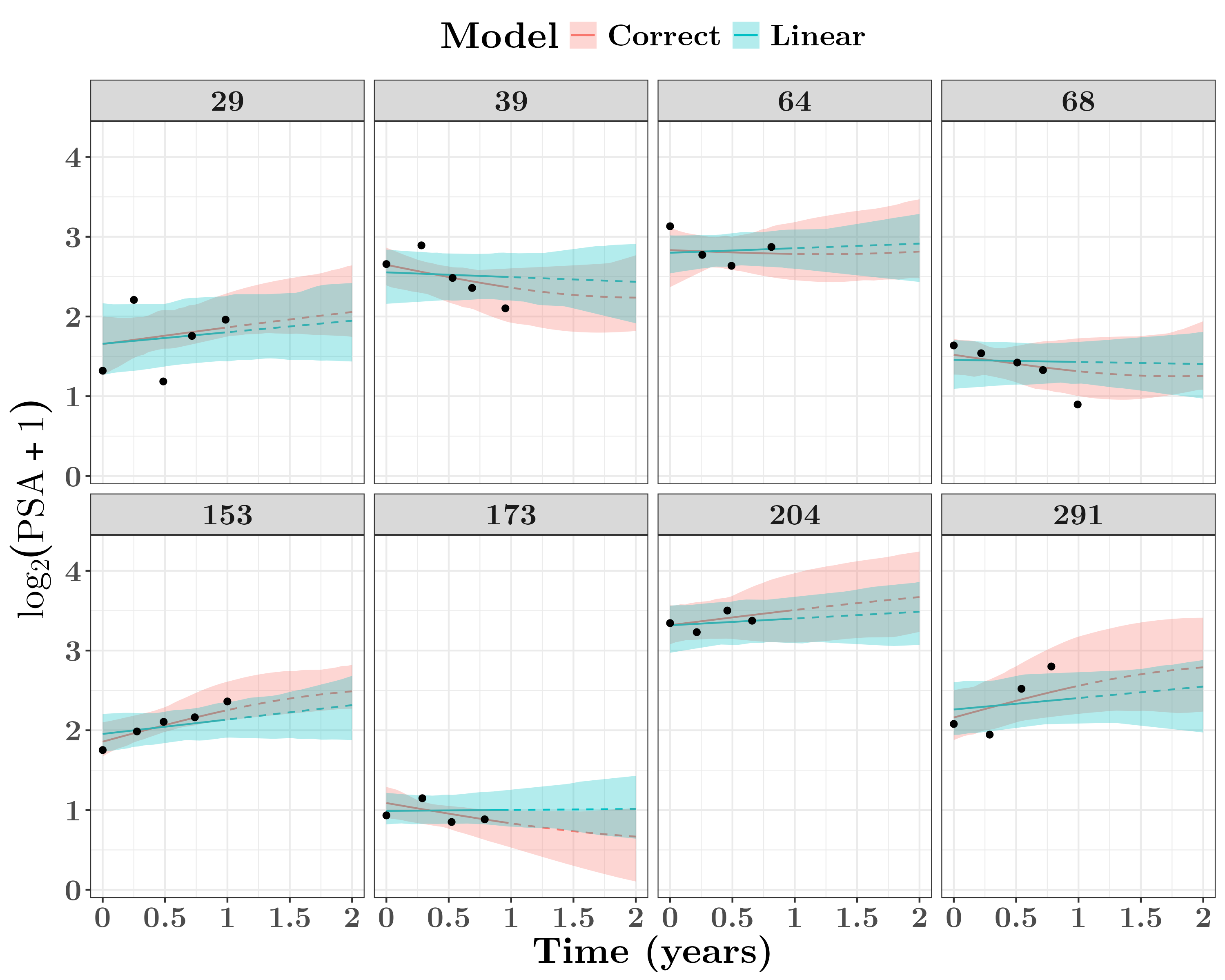}
\caption{The observed trajectories (in black points), fitted trajectories (with 95\% credible interval presented in ribbons) assuming a linear effect of time (in blue, as in the linear models) and natural cubic splines with 3 degrees of freedom (in red, as in the correctly-specified models) of 8 randomly selected patients from the simulated datasets. The fit for the time after year 1 (indicated by dashed lines) is included to provide a better picture of the degree of non-linearity of the spline fit.}
\label{fig:psatra}
\end{figure}

We randomly select 8 patients from the first evaluation (on the second simulated dataset) and show the fitted trajectory of PSA (transformed in $\log_2(\text{PSA}+1)$) resulting from the models assuming a linear trajectory versus natural cubic splines (with 3 degrees of freedom, i.e., the correctly-specified model). The trajectories are shown in Figure~\ref{fig:psatra}.

\subsection{Calculation of EPCE in the scenario without censoring}

In the manuscript, we showed the estimation of the EPCE using the model-based risks, in the setting with interval censoring and competing risks. For the simulation study, the reference EPCE is calculated with the true event times, thus without interval cenosring. It can be estimated by
\begin{align*}
\widehat{\text{EPCE}}^{\textsc{prg}}_{\text{ref}}(t+\Delta t, t) = \frac{1}{n_t}\sum_{i: \min(T^{\textsc{prg*}}_i, T^{\textsc{trt*}}_i) \geq t}-\log\left[p\{\tilde{T}_i, \tilde{\delta}^{\text{ref}(1)}_i, \tilde{\delta}^{\text{ref}(2)}_i \mid T^{\textsc{prg}*}_i \geq t, T^{\textsc{prg}*}_i < T^{\textsc{trt}*}_i, \boldsymbol{\mathcal{Y}}_i(t), \boldsymbol{\mathcal{D}}_n\}\right],
\end{align*}
where $\tilde{T}_i = \min(T^\textsc{prg*}_i, t + \Delta t)$, 
$\tilde{\delta}^{\text{ref}(1)}_i = I\left(T^{\textsc{prg*}}_i \geq t, T^{\textsc{prg*}}_i < t +\Delta t,T^{\textsc{prg*}}_i < T^{\textsc{trt*}}_i\right)$, and $\tilde{\delta}^{\text{ref}(2)}_i = I\left\{T^{\textsc{prg*}}_i \geq \min(T^{\textsc{trt*}}_i , t+ \Delta t)\right\}$. The term in the log function can be estimated by
\begin{align*}
p\{\tilde{T}_i, \tilde{\delta}^{\text{ref}(1)}_i, \tilde{\delta}^{\text{ref}(2)}_i \mid T^{\textsc{prg}*}_i \geq t, T^{\textsc{prg}*}_i < T^{\textsc{trt}*}_i, \boldsymbol{\mathcal{Y}}_i(t), \boldsymbol{\mathcal{D}}_n\} = \log\left[\tilde{\delta}^{(1)}_i \mathcal{F}^{\text{ref}}_1 + \tilde{\delta}^{(2)}_i \mathcal{F}^\text{ref}_2 \right],
\end{align*}
where the first factor $\mathcal{F}^\text{ref}_1 = \Pr\{T^{\textsc{prg}*}_i \leq \tilde{T}_i \mid T^{\textsc{prg}*}_i \geq t, T^{\textsc{prg}*}_i < T^{\textsc{trt}*}_i, \boldsymbol{\mathcal{Y}}_i(t), \boldsymbol{\mathcal{D}}_n\}$ is the cumulative incidence function of progression until $\tilde{T}_i$, and the second factor $\mathcal{F}^\text{ref}_2 = \frac{\Pr\{T^*_i \geq \tilde{T}_i \mid \boldsymbol{\mathcal{Y}}_i(t), \boldsymbol{\mathcal{D}}_n\}}{\Pr\{T^{*}_i \geq t \mid \boldsymbol{\mathcal{Y}}_i(t), \boldsymbol{\mathcal{D}}_n\}}$ is the overall survival probability of the patient surviving after $\tilde{T}_i$ conditional on that he did not experience either of the events until $t$.

\bibliographystyle{plainnat}
\bibliography{ref}